\documentclass[
preprint, aps]{revtex4}
\usepackage{graphics}
\usepackage{graphicx}
\usepackage{fancybox}

\begin{document}


\title{Multiband Hubbard Models and the Transition Metals}

\author{Gernot Stollhoff}

\affiliation{
Max-Planck-Institut f\"ur Festk\"orperforschung               \\
Heisenbergstra\ss e 1,  D-70569 Stuttgart, Germany            \\
}

\begin{abstract}
Correlation computations on multiband Hubbard Hamiltonians
are presented. It is shown why the proper degeneracy is of vital
importance and that the atomic exchange interaction plays a
particular role. The different methods are connected, and their results
are discussed. Many experimental properties
for the elemental solids can be explained by single closely related
sets of parameters each.
There is an exception, $T_c$ of $Fe$. Here, a
novel feature of longer-range interactions enters. Connections are
made to LSDA-calculations; and their seeming successes and deficiencies
are explained.  

\end{abstract}
\maketitle

\tableofcontents

\section{  Introduction    }             \label{sec:intro}
  
A breakthrough in the understanding of the itinerant ferromagnetism of the
$3d$-transition metals occured when density functional (DF) calculations,
performed in the local density approximation (LDA) \cite{LDA1,LDA2}
or rather its
spin generalized variant, local spin density approximation (LSDA) \cite{6},
obtained seemingly perfect values for the magnetic moments of the metals. 
Good results
were also obtained for other ground state properties: binding energies, 
equilibrium volumes, bulk moduli, and Fermi surfaces, to name a few
\cite{15,16,17}.

A crucial aspect in this context is the behaviour of the $3d$-electrons. 
The $3d$-atomic orbitals are largely localized
in space around the nuclei, and form rather narrow tight binding bands 
with a width of roughly 5eV. This localization is considerably stronger for 
the $3d$ than for the heavier $4d$ or $5d$-transition metals.  
In the following work, the consequences of this localization are 
investigated for the $3d$-transition metals.

While there have been many speculations about a full localization of the
$3d$-electrons and about a representation of their degrees of freedom by
spin-Hamiltonians, the delocalization is a feature
that is strongly supported by experiment. Already before the success of LSDA, 
the experimental binding energies were unambiguously connected to 
delocalized $3d$-electrons \cite{38}.
In LSDA, the explicit part of the
kinetic energy is obtained from a maximally delocalized single-determinant
reference state while correlation corrections are contained in a functional.
Consequently the former part represents the limit of
maximal delocalization of the $3d$-electrons. On the 
other hand there is evidence that the LSDA results  
overemphasize binding by up to 20 percent \cite{18}, possibly due to
a mishandling  of correlation corrections. 
This deviation in turn sets an upper limit
to possible correlation corrections.  In particular strong 
correlations or electronic degrees of freedom that can be described in terms of
atomic moments are ruled out.

With the $3d$-electrons basically delocalized in 
tight-binding bands, magnetism must result from the electronic interactions.
As already mentioned, these cannot be so strong that localized
atomic moments arise. Even more, it needs to be understood why
the LSDA is apparently able to handle these interactions that are
definitely connected to the spatially strongly localized atomic orbital 
representation of the $3d$-elctrons. It is hard to imagine how a 
homogeneous-electron gas exchange-correlation potential can adequately deal
with such inhomogeneous atomic features. In fact it cannot.

Bare local Coulomb-interaction
matrix elements  between electrons in these atomic orbitals are of the
order of 20eV, far larger than the kinetic energy gained
from delocalization into bands. Therefore, a second problem is to 
understand how these bare interactions are reduced into
the required smaller interactions.

The optimal way to deal with these questions would be a full ab-initio 
correlation calculation. The only method available for such a treatment 
is the Local Ansatz (LA) \cite{stol3,stoln}. It starts from Hartree-Fock
(HF) ab-initio calculations for solids and adds correlations in a variational
way like Quantum Chemistry (QC) methods do. However, it differs from
these methods by using specifically constructed subsets of correlation
operators with a well defined local meaning instead of trying to
cover the whole correlation space in an orthogonal representation.
This results in a loss of typically one to a
few procent of the correlation energy in a given basis, but it leads to a 
large 
gain in efficiency and it enables the LA to treat metals. 
First ab-initio calculations for metals \cite{li} and for a metallic 
transition 
metal compound \cite{s98}  have already been performed, and 
calculations for
non-magnetic transition metals are under way \cite{sun};
however, a complete coverage of the
magnetic problem has not yet been obtained. 
Therefore, the problems mentioned above could so far only be 
addressed using correlation calculations for  models. 

The minimal level of complexity for such models is well defined:
a tight-binding Hamiltonian for the $3d$-electrons,
i.e. a five-band (per spin) Hamiltonian. The hope is that
the $4s$- and $4p$-orbitals of the transition metals do not need to be
explicitly included for the basic understanding of magnetism, 
since they contribute little to
the electronic density of states in the relevant energy range around the 
Fermi energy. This omission certainly causes defects, for example
of the Fermi surface.
For the  interactions, a first choice is the inclusion of only 
local (atomic) interactions of the $3d$-electrons. It is known that these 
can be condensed into
three Slater parameters. We will in the following rearrange 
those  terms and call the resulting interactions Hubbard ($U$) interaction,
Hunds rule exchange terms ($J$), and anisotropy terms ($\Delta J$) \cite{11}.
The underlying assumption is that longer range contributions
of the Coulomb interaction are almost perfectly screened for these metals.

Such models have for long been the basis of attempts to understand
the itinerant ferromagnetism. 
However, these attempts were mostly restricted to simplified single-band 
models and/or
to the approximate treatment of the interaction in Hartree-Fock (HF)
approximation,
or when the treatment was extended to finite temperatures
within a functional integral formulation, 
to an equivalent static approximation. 
For an early overview, I refer to \cite{mor1}. 

Here, the LA led to a sizeable improvement.
Since it can be applied to models as well as to ab-initio calculations,
we were able to perform satisfactory correlation calculations
for the model described above, and we have computed
the non-magnetic \cite{st} as well as the magnetic \cite{os} cases.
The tool to understand the magnetic phase transition for the
case of delocalized electrons is the Stoner-Wohlfarth 
theory \cite{sto,wo}. Such an analysis had been earlier performed in the case
of  LSDA computations \cite{gunn1}. We also analysed our
results in the same way \cite{os85,soh}, and managed to
work out why the LSDA calculations had been so successful for the transition
metals but had failed for a set of transition metal compounds \cite{soh,soh1}.
We abstained from any attempts to generalize the treatment of the order
parameter beyound a mean-field (or Stoner-Wohlfarth) approximation.
It should be noted that for the simplified handling of the interaction
in HF or static approximation, a generalized spin fluctuation theory
is available \cite{mor2}. 

It turned out from our analysis that the 5-fold degeneracy of the 
model bands is very
relevant. Single-band or two-band models are not able to catch the
essence of the $3d$-magnetism at all. Many of the degeneracy features also
are lost when restricting to a HF or static approximation.
As will be shown below, a reduction of the degeneracy would require
also larger and larger interactions and would incorrectly push the 
treatment into
a strong-correlation direction
which is inadequate for the $3d$-transition metals.

Five band models have been rarely treated by other methods beyound 
HF-approximation. A first attempt
was made in the context of an insufficient second-order perturbation
computation \cite{35,36}. Quasiparticle calculations followed using the
Kanamori t-matrix approach \cite{62} for almost filled 
degenerate band
systems such as  $Ni$ \cite{58,59,60,61}. 
Recently, calculations
have been performed for nine-band models starting from an $R=0$-approximation
that had also been used with the Local Ansatz but employing a full 
Configuration Interaction (CI) calculation instead of the weak-correlation
expansion or of a two-particle excitation CI calculation both within
the LA \cite{web,web1}. Finally, the Dynamical Mean-Field-Theory (DMFT) 
has been used
for such a model \cite{li2}. The latter is the appropriate generalization
of the functional integral schemes just mentioned, and goes beyound the
static approximation.

We will in the following introduce the five-band model plus its interactions,
and describe in detail the LA treatment and the different approximations made.
We will also establish connections to the other computational schemes.

In a next step,
results of the  calculations for the non-magnetic case will be
presented, and the different approximations will be tested.
A further step is to compute and analyse the magnetic results
for $Fe$, $Co$, and $Ni$.
Based on them, conclusions with regarding the comparability to experiment 
and assessing the
specific deficiencies of the LSDA and its results will be made.

Finally, connections between the model and first ab-initio correlation
results will be made, and the limits of the 
Hubbard-model scenario will be revealed.

     \section{Model Hamiltonian and Single-Particle Groundstate}     
\label{sec:model}

The aim of the qualitative treatment is to understand the
delocalization and interaction of the $3d$-electrons which is 
the expected key for the understanding of magnetism in the
3d-elemental solids. There are five $3d$ electrons per spin and site
 (atom). A compact description of their delocalization is in the form of
canonical d-bands \cite{jep}. This is essentially a tight-binding
description and has the additional advantage of containing only a single
open parameter, namely the $3d$-band width $W$. The single-particle
part of the model Hamiltonian ($H_0$) is given in terms of these orbitals in
eigenvalue representation. In the computations, these bands are constructed
for the two relevant lattices, bcc and fcc.

 For these electrons we further assume that they only
interact when they are on the same atom $l$. These interactions can be
given in terms of three Slater interactions; here we choose a slightly 
different but equivalent notation. The full Hamiltonian $H$ reads
\begin{eqnarray}
H & = & H_0 + H_1 \\
H_0 & = & \sum_{\nu\sigma{\bf k}} e_{\nu }({\bf k}) n_{\nu \sigma}({\bf k})\\
H_1 & =& \sum _{\bf l}H_1({\bf l}) \\
H_1({\bf l})&=&{1\over 2}\sum_{ij\sigma \sigma ´}
U_{ij}a_{i\sigma}^\dagger (l)a_{j\sigma '}^\dagger (l)a_{j\sigma '}(l)
a_{i\sigma}(l)\\
&&+{ 1\over 2}\sum_{ij\sigma \sigma ´}
J_{ij}[a_{i\sigma}^\dagger (l)a_{j\sigma '}^\dagger (l)a_{i\sigma '}(l)
a_{j\sigma}(l)+a_{i\sigma}^\dagger (l)a_{i\sigma '}^\dagger (l)a_{j\sigma '}(l)
a_{j\sigma}(l)].
\label{eq12}
\end{eqnarray}

The  $e_{\nu }({\bf k})$ represent the five ($\nu =1..5$)
canonical bands, and the $n_{\nu \sigma}({\bf k})$ the corresponding number 
operators of the Bloch eigenstates, whose  creation and 
annihilation operators are $c_{\nu \sigma}^\dagger ({\bf k}),
c_{\nu \sigma} ({\bf k})$.
 The $U_{ij}$ and $J_{ij}$ are the
local (atomic)interaction matrices and related by
\begin{equation}
U_{ij}=U+2J-2J_{ij},
\end{equation}
where $U$ and $J$ are the average Coulomb and exchange interaction constants.
The matrix $J_{ij}$ contains the third interaction parameter $\Delta J$ that
is a measure of the difference between the $e_g$ and $t_{2g}$ interactions.
For details of this matrix, we refer to \cite{11,os}. For $\Delta J=0$,
it holds that $J_{ij}=J$. The interactions are expressed in terms of the
five 
local $3d$-orbitals $i$ on atomic positions $l$
whose creation and annihilation operators are given as
$a_{i\sigma}^\dagger (l)a_{i\sigma}(l)$.
 
The size of these parameters will be fixed later.
Typically, it holds that the band width $W$ that scales
the single-particle part $H_0$ is roughly 5eV. The interactions
are reduced to a single free parameter by setting $J\simeq0.2U$
and  $\Delta J \simeq 0.2J$. For $U$ it holds that typically
$U \simeq 0.5W$.  

Starting point of the following correlation treatment is the solution to the 
single-particle Hamiltonian $H_0$, called $\Psi_{0}$. This is written as
\begin{equation}
|\Psi_{0}(n_d,0)\rangle = \prod_{{{\bf k}\nu \sigma} 
\atop {e_{{\bf k}\nu}\leq e_F (n_d)}}
c_{\nu \sigma}^\dagger ({\bf k})|0\rangle.
\end{equation}

This solution differs slightly from the self-consistent field (SCF)
solution of the full model Hamiltonian $H$. The latter generates
additional selfconsistently obtained crystal field terms
that may lead to charge redistributions between the $e_g$ and $t_{2g}$
orbitals. However, since the interactions are not too large, and the
original site occupations are almost degenerate, these  redistributions
can be almost neglected. An exception that will be discussed later is the
case of ferromagnetic $Ni$ where the second requirenment does not hold.
Such a solution is found for all fillings $n_d$ per atom of the 
five-band system, with
$n_d$ ranging from 0 to 10. $e_F(n_d)$ is the occupation 
dependent Fermi energy.

In addition to this non-magnetic solution, magnetic solutions
with a moment $m$ are constructed by generating states
\begin{equation}
|\Psi_{0}(n_d,m)\rangle = 
\prod_{{{\bf k}\nu} \atop {e_{{\bf k}\nu}\leq e_F(n_d+m)}}
\prod_{{{\bf k'}\nu} \atop {e_{{\bf k'}\nu}\leq e_F(n_d-m)}}
c_{\nu \uparrow}^\dagger ({\bf k})c_{\nu \downarrow}^\dagger ({\bf k'})
|0\rangle.
\end{equation}

Here, the majority band with spin up is occupied with $ {n_d+m \over2} $
electrons, and the minority band with spin down with   $ {n_d-m \over2} $.
This ansatz is a rigid band approach. Again, a self-consistent
solution might lead to small charge redistributions in the minority
and majority bands if these are not empty or filled, respectively.

The Fermi energy for the individual cases like $Fe, Co, Ni$ is chosen so that
the maximal magnetic moment agrees with the same one of the LSDA calculations.
This implies an occupation of 7.4, 8.4, 9.4 for $Fe, Co$ and $Ni$, 
respectively. It is known from more careful charge analyses that the true
atomic $3d$-occupations are somewhat smaller. For $Fe$ for example they amount
to 6.5. Consequently, the $d$-orbitals of this model Hamiltonian are
not maximally localized $3d$-tight binding orbitals but their tails have 
small $4s$-, $4p$-contributions.

     \section{Correlated Ground State}     
\label{sec:corr}

\subsection{Deficiencies of the Single-Particle Ground State}
The single-particle ground state $\Psi_0$ is an eigenstate of
the single-particle
part of the Hamiltonian but results in a poor coverage of  the interaction 
part. 
Being represented by eigenfunctions in momentum space, this state has
maximal local charge fluctuations that are uncorrelated for
the different bands. A measure of these charge fluctuations is the
atomic quantity $\Delta n^2$ for a given wave function. It is given as
\begin{eqnarray}
\Delta n^2 &= &\langle \Psi |n^2(l)|\Psi \rangle -\langle \Psi |n(l)|\Psi \rangle ^2 \\
n(l)&=&\sum_{i \sigma} n_{i \sigma} (l),
\end{eqnarray}
where $n_{i\sigma}(l)$ is the density operator for an electron with 
spin $\sigma$ in orbital $i$ on site $l$.

For the single-particle ground state, it holds that
\begin{eqnarray}
\Delta n^2 (\Psi_0(n_d,0))&= &\sum_{i \sigma} n_{i \sigma}(1-n_{i \sigma})\\
n_{i\sigma }&=& \langle \Psi_0 | n_{i \sigma}(l) | \Psi_0 \rangle .
\end{eqnarray}
These fluctuations increase linear with the number of bands.
For the half-filled five-band case with degenerate occupation we find that
$\Delta n^2 =2.5$. Correspondingly, the interaction energy costs per atom 
of this state in comparison to the disordered atomic limit
amount to $1.25 (U-{2 \over 9}J)$. This needs to be compared to a kinetic
energy gain which, for a roughly constant density of state, equals  $-1.25W$.
This indicates that half the delocalization energy is lost in this
approximation for a ratio $U \simeq 0.5W$. 
However, the electrons on the individual atoms
order by Hund's rule and can gain an energy of $-{70 \over 9}J $
at half-filling.
Consequently, even for this relatively modest screened interaction, the solid
is no longer bound in single-particle approximation and a better 
treatment is required, i.e. the correlated ground state needs to be computed.
Without including correlations explicitly, any broken symmetry, 
even disordered local moments, would be favourable.

\subsection{The Local Ansatz}

For a three-dimensional model with five degenerate bands,
the correlation treatment can not be done exactly
but only approximately. Since the parameter choice lets us
expect that the electrons are not too strongly correlated, the natural
approach is to start from the single-particle ground state and add 
correlations as corrections.

This is how the Local Ansatz is set up. Here,
the following variational ansatz is made for the correlated ground
state:
\begin{eqnarray}
| \Psi_{corr} \rangle & = & e^{-S} | \Psi_{0} \rangle  \label{eq:lanst0}   \\
      S         & = &  \sum_{\nu} \eta_{\nu} O_{\nu}    \\
O_{\nu}                   & = &  \left\{ \matrix{
                                      n_{i \uparrow} (l)n_{i \downarrow}(l) \cr
                                      n_i (l)n_j(l')              \cr
                                      \vec{s}_i (l)\cdot \vec{s}_j (l')  \cr} 
   \right.
\label{eq:lanst1}
\end{eqnarray}
The $n_{i \sigma}(l)$
and $\vec{s}_i(l)$ are density and spin operators for an electron in the
local orbital $i$ on site $l$.
The operators have a straightforward interpretation. For example, the 
first operator $n_{i \uparrow} (l)n_{i \downarrow}(l)$, 
when applied to $ | \Psi_{0} \rangle $,
picks out all configurations
with two electrons in orbital $i$.
When applied with a variational
parameter $ \eta_{\nu}$,
as in eq.(\ref{eq:lanst0}), it partially suppresses those
configurations. For a single-band Hubbard model, such an ansatz was first
made by Gutzwiller \cite{33}.
Similarly, the operators
$ n_i(l) n_j(l') $
introduce density
 correlations between electrons in local orbitals $i,j$ either on the same 
site or on different sites $l,l'$.
The wave function generated by these two sets of operators, when applied to
the homogeneous
 electron gas problem, is the
Jastrow function \cite{jastr}.
The operators
$ \vec{s}_i \cdot \vec{s}_j $
generate spin correlations. On the same site, they introduce Hund's
rule correlations, while when applied for different sites they 
result in magnetic correlations. 

For the same sites, all these operators are directly connected
to $H_1$. They allow to correct exactly those features that
are addressed by the interacton terms. 

In the following, we will no longer use the full
operators but only their two-particle excitation 
contributions.
The standard approximation to derive the energy and to obtain the variational
parameters is an expansion in powers of $\eta$, up to second
order,
\begin{eqnarray}
E_G      & = & 
       E_{SCF} + E_{corr}                 \\
E_{corr} & =  &  - 2\sum_{\nu} \eta_{\nu}
\langle O^{\dagger}_{\nu} H\rangle
     + \sum_{\mu \nu}\eta_{\nu} \eta_{\mu} \langle O^{\dagger}_{\nu}
HO_{\mu}\rangle_c  \ \ . \label{eq:expan0}
\end{eqnarray}
When optimizing this energy, the following equations arise that determine
the energy and the variational parameters.
\begin{eqnarray}
E_{\rm corr} & = &  - \sum_{\nu} \eta_{\nu} \langle 
O^{\dagger}_{\nu} H\rangle \label{eq:expan1} \\
     0 & =  &  - \langle O^{\dagger}_{\nu} H\rangle
     + \sum_{\mu} \eta_{\mu} \langle O^{\dagger}_{\nu}
HO_{\mu}\rangle_c
     \ \ .               \label{eq:expan2}
\end{eqnarray}
Here, $\langle A \rangle$ is the expectation value of the operator
$A$ within
$| \Psi_{0}\rangle$, and $\langle \rangle _c$ indicates that only connected 
contributions are included. It holds that $E_{SCF}=\langle H \rangle$
These equations  can also be identified as a Linearized
Coupled Cluster expansion, restricted to particular two-particle
(double) and one-particle (single) excitations, abbreviated LCCSD.
The concept of CC-equations was introduced into many-body
physics and into Quantum Chemistry a long time ago \cite{cc1,cc2,cc3,cc4}.
In the presented calculations on models,
single-particle excitations were,
in contrast to the ab-initio calculations,  not included in the
correlation treatment itself,
but were covered by direct modifications in the trial single-particle
wave functions.

The full treatment of these equations poses no problems, in particular when
the operators are restricted to on-site terms. Then, without consideration
of intrinsic symmetries, one has 5 Gutzwiller-like operators and 10
density and spin operators each. The most expensive step is the solution of a
set of linear equations with dimension 25. 

\subsection{The $R=0$-Approximation and Alternatives to the Local Ansatz}

The  correlation calculations can be performed in 
a further approximation in which one may go  beyond the LCCSD-equations.
This approximation applies only when restricting to on-site operators.
In this case one can approximately set all those terms in
the matrices $\langle OH \rangle$ and $\langle OHO \rangle$ equal to 
zero where the operators in the $O$ or in $H_1$ are not on the same site. 
In this approximation, the full correlation treatment separates into
independent contributions covering a single site each in an non-correlated and 
non-interacting environment.
 This approximation is called single-site or $R=0$-approximation.
It was introduced for the first time in a
second-order perturbation treatment of a five-band Hamiltonian that is 
very similar to the one used here. There, however, no restriction
to a particular choice of correlation operators was made but the full 
two-particle operator space was covered \cite{36}. 
In the single-site approximation of the LA,
all required terms are simply obtained from two sets of single-particle
elements, the individual occupations $n_{i\sigma}$ and the average
energies on these sites, $e_{i\sigma}$, given as
\begin{equation}
e_{i\sigma }=\int_{-\infty }^{e_F}en_{i\sigma}(e)de
\end{equation} 
where $n_{i\sigma}(e)$ is the local partial density of state for orbital $i$
with spin $\sigma$.
More details can be found in refs. \cite{st,os}. All terms that
are left out in the single site approximation contain non diagonal density
matrix elements of the form 
$P_{ij}(l,l')=\langle a_i^{\dagger}(l)a_j(l')\rangle$ with $R=l-l'\neq 0$,
explainng the name.
With rising number
of nearest neighbors, these contributions shrink in weight and disappear
for the limit of infinitely many neighbors (or equivalently dimensions $d$).
The approximation was therefore more recently called the
$d=\infty$-approximation \cite{vo}.

In a way, this approximation represents the correlation generalization 
of  an approximate coherent potential approximation (CPA) where
a single-site mean-field calculation with broken symmetry is performed which 
leads to disordered local moments.
With on-site correlations properly included, a broken symmetry
result does no longer arise, at least not prior to a Mott-Hubbard
transition. Or, in different words, disordered local moments are
a poor man's approach to correlations.

This single-site approximation allows a more general treatment than
the  LCCSD-approximation. One possibility is to perform a $CI$-calculation
for this single site. The exact energy of the variational state
with two-particle excitations included is obtained. It represents a lower limit
to the exact result, while the LCCSD-approximation usually overshoots the
latter. The correlated wave function for the single site $l$ is defined 
as
\begin{equation}
| \Psi_{corr} (l)\rangle  = (1-\sum_{\nu \in l}O_{\nu})| \Psi_{0} 
\rangle . \label{eq:ci0}  
\end{equation}

Its exact energy is
\begin{eqnarray}
E_G (l)      & = & 
       E_{SCF} + E_{corr} (l)                 \\
E_{corr} (l)& =  & { {- 2\sum_{\nu}' \eta_{\nu}
\langle O^{\dagger}_{\nu} H'\rangle
     + \sum_{\mu \nu}'\eta_{\nu} \eta_{\mu} \langle O^{\dagger}_{\nu}
H'O_{\mu}\rangle_c}
\over {1+ \sum_{\mu \nu}'\eta_{\nu} \eta_{\mu} \langle O^{\dagger}_{\nu}
O_{\mu}\rangle_c }} \ \ .\label{eq:ci1}
\end{eqnarray}
Here, the $\sum'$ indicates a restriction of the 
summation to operators on site $l$,
and $H'=H_0+H_1(l)$.
When optimizing this energy, the resulting equations that determine
the energy and the variational parameters can be written in close similarity
to the LCCSD-equations \ref{eq:expan1},\ref{eq:expan2}.
\begin{eqnarray}
E_{\rm corr}(l) & = &  - \sum_{\nu}' \eta_{\nu} \langle 
O^{\dagger}_{\nu} H'\rangle \label{eq:ci2} \\
     0 & =  &  - \langle O^{\dagger}_{\nu} H'\rangle
     + \sum_{\mu}' \eta_{\mu} \langle O^{\dagger}_{\nu}
H'O_{\mu}\rangle_c \nonumber \\ 
 & &    + \sum_{\mu}' \eta_{\mu} \langle O^{\dagger}_{\nu}O_{\mu}\rangle_c
     \sum_{\mu'}' \eta_{\mu'} \langle
H'O_{\mu'}\rangle_c
     \ \ .               \label{eq:ci3}
\end{eqnarray}
The newly added terms in eq. \ref{eq:ci3} are responsible for the difference.
One may generalize in this approximation and
 perform a full $CI$-calculation not restricted to two-particle
operators. The system of equations 
(\ref{eq:ci0},\ref{eq:ci1},\ref{eq:ci2},\ref{eq:ci3}) 
stays the same, but the operators $O_{\nu}$ are not
restricted to two-particle excitations.
The available operator space explodes exponentially 
with degeneracy, and the numerical demands rise sharply, but recently
even nine-band models could be addressed this way 
(i.e., the $3d$- plus the $4s$- and $4p$-orbitals 
were included)\cite{web,web1}.

An even more extended approach is to perform an exact calculation for
the single site problem, based on dynamically fluctuating disordered
local moments, called DMFT (for an introduction of its origins, see ref.
\cite{vo}). This computation is
based on a Green's function formalism, and is the by far most expensive
method. Formally, the operator space is extended beyond strictly
local operators. Not all electrons are covered equally but their
treatment is influenced by their individual energies.
Applications to a nine-band model were done for
$Fe$ and $Ni$ \cite{li2} using this method. A further 
advantage of this scheme is that 
 quasiparticle results can
be obtained and a transition to thermodynamic quantities is possible,  
since the computations are performed on a Green's function level.

Let's return to the most simple scheme, the LA. In contrast to
the other methods, it cannot be applied to the Mott-Hubbard transition.
However, it is the 
only scheme that can be extended beyond $R=0$ and can treat finite 
dimension corrections. Furthermore, it can deal with
long range interactions, and even manage ab initio calculations
with the full interaction.

There is actually a rather simple extension of the LA that makes it
more tolerant of stronger correlations. This is the application
of the full CCSD-equations. They arise from a full equation-of-motion
of the original ansatz for the wave function, and not from a weak-coupling
approximation. The resulting equations  read as
\begin{eqnarray}
E_{\rm corr} & = &  - \sum_{\nu} \eta_{\nu} \langle 
O^{\dagger}_{\nu} H\rangle \label{eq:cc2}\\
     0 & =  &  - \langle O^{\dagger}_{\nu} H\rangle
     + \sum_{\mu} \eta_{\mu} \langle O^{\dagger}_{\nu}
HO_{\mu}\rangle_c \nonumber \\&&
     - {1\over 2}\sum_{\mu\mu'} \eta_{\mu}\eta_{\mu'}\langle O^{\dagger}_{\nu}H
O_{\mu}O_{\mu'}\rangle_c
     \ \ .               \label{eq:cc3}
\end{eqnarray}
This
generalization would improve the results for larger ratios of ${U \over W}$.
When applied to a single-band Hamiltonian in the $R=0$-approximation, the
result should, with single-particle operators properly added, reproduce
the Gutzwiller approximation \cite{stohei}.

\subsection{Error Estimates of the different Approximations in the 
Local Ansatz}

In the present treatment of the LA, three approximations 
are made that need to be controlled.

The first is the weak-correlation approximation. It may be tested
by a comparison of the LCCSD- and the CI-result in the $R=0$-approximation.

The second is the $R=0$-approximation itself. It can be tested only in 
the weak-correlation approximation.

The last is the restriction in correlation operator space. Again,
this will be tested in the weak correlation expansion.

The last restriction is best tested for the one-dimensional single-band
Hubbard model. Here, the exact energy is known, and also its weak-correlation
limit. The corresponding Gutzwiller ansatz which contains only 
single-site operators yields 92 percent of
the exact correlation energy in this limit for the half-filled case. 
A large fraction of the missing energy can be obtained by longer-range
density and spin correlations. 

The situation is worse for almost empty bands. 
Here, additional correlation operators of the form $[H_0,O_{\nu}]$ are
required for a satisfactory result. Such operators do not open new correlation
channels but allow us to take into account the band energies of the electrons
involved in the correlation process. Such operators are  included in the
ab initio scheme and have turned out to be important in a different context 
\cite{stoln}. We have found for the non-magnetic 
five-band calculations \cite{st,s85} that these operators 
do not lead to noticable changes. Although being of non-local nature, such
operators contribute in the $R=0$-approximation and bridge the 
difference between
a correlation calculation restricted to local correlation operators and
a DMFT calculation.

For the five-band model,
we had also included nearest neighbor operators in the non-magnetic
calculations \cite{st}. The energy gain due to these terms was only a few 
percent; thus we can trust the results of calculations restricted to 
on-site correlations for the five-band Hamiltonian.

The $R=0$-approximation depends on the number of neighbors. For a two-site
problem with a single orbital each, the $R=0$-result is only half the correct 
result. For the one-dimensional Hubbard model, the $R=0$ result needs
to be enhanced by 33 percent to obtain the final LA result with on-site
operators for half-filling, but already nearest neighbor or $R=1$-corrections 
reduce the deficiency to one percent. For the five-band
problems treated here, the $R=1$-corrections turned out to be 2-3 percent for
fcc or bcc, respectively \cite{st}. Most of our calculations and all calculations
for the magnetic state were therefore restricted to the $R=0$-approximation.
On the other hand, 
these results indicate that an $R=0$ or 
$d_{\infty}$-approximation should not be applied to systems with less 
than six neighbors. 
\begin{figure}[h,b,t]
{\centerline{\includegraphics[width=8cm,clip,angle=0]{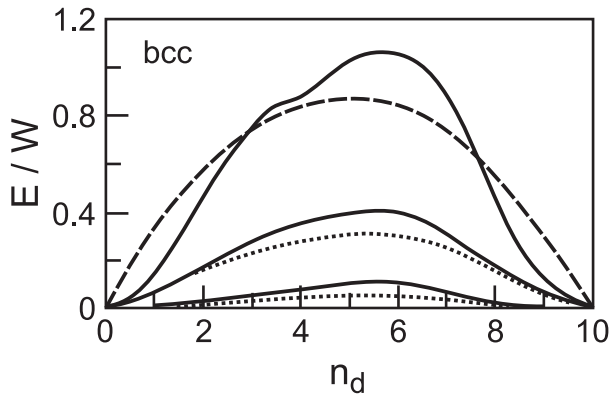}}}
\caption{Several energies in units of $W$ as a function of 
band filling $n_d$. The interaction energy costs in SCF-approximation
are given by the dashed line. The top full line gives the MP2 correlation
result, the second full line the LCCSD-result, and the dotted line just
below it the CI-result. The lowest full line and the dotted line below
give the amount of correlation energy that is lost when the spin correlation
operators are omitted from the LCCSD or CI calculations, respectively.
All correlation calculations were performed for the bcc case and in 
the $R=0$-approximation (from ref. \cite{st}).}
\label{fig1}
\end{figure}

Let us finally turn to the validity of the weak-correlation approximation.
It will definitely fail around half-filling for ratios ${U \over W} > 1$.
In Fig. \ref{fig1}, correlation energies in the $R=0$-approximation are
given for a ratio of $U=0.5W$. The dashed curve gives the $HF$-energy costs
$\Delta E_{SCF}=E_{SCF}-\langle H_0 \rangle$.
The topmost curve gives the result of a correlation calculation
performed in second-order perturbation (MP2) expansion (the LCCSD-equation
can be reduced to MP2 by replacing
$\langle OHO\rangle$ by $\langle OH_0O\rangle$). This result overshoots the
HF-term and is definitely wrong. The second solid curve gives the final result
in LCCSD, and the dotted curve below it gives the CI-result. The maximal
relative difference around half-filling amounts to 25 percent. When 
considering that the CI-result is a true lower limit, then the LCCSD-result
should not overshoot the correct result by more than 5 percent.
The lowest curves display the energies that are lost when spin-correlations 
are omitted.
Here, the relative differences between CI and LCCSD are considerably larger 
but again the true result is expected to be close to the LCCSD result.

These particular features are closely connected to the degeneracy.
As mentioned before, fluctuation costs arise in five channels $i$ in
the HF-approximation. The correlation ansatz, however, makes
15 density correlation channels $i,j$ available. Thus, 
treating these channels independently as is done in perturbation theory,
very soon overscreens the fluctuations. The term $H_1$
in $\langle OHO \rangle$ by which the
LCCSD-equations differ from MP2 guarantees that the different 
channels take note
of each other and act coherently. This distribution of correlation corrections
among many different states also makes it plausible why the CI and 
LCCSD results 
are so close to each other
although five degenerate orbitals need to be treated. 

For the spin correlations,
the situation is different. Here each pair $i,j$ can gain an interaction energy
$J$ independent of each other. This is reproduced in the LCCSD-equations,
while it is a particular feature of the CI-calculations restricted to
two-particle excitations
that at each moment
one has either the one or the other electron pair corrected. Thus, the 
different contributions, also the spin and density contributions
 actually impede each other, and a considerably smaller energy
is obtained. This explains why
the largest part of the difference between the two schemes
in the full calculation arises from the addition
of the spin correlations. On a CI-level,
this deficiency might only be corrected by including in the variational
ansatz for the CI-wave function
not only two-particle excitations but also their products, and finally
up to ten-particle excitations.

The failure of MP2 found here is related to the failure of this
approximation when the screening of the long-range Coulomb interactions
in metals is concerned. There, fluctuations are also 
diagonal, i.e. connected to the
local sites $l$. The correlation space, however, offers $l,l'$ density 
correlations that are independent in MP2 and cause the well-known divergence 
of the correlation energy. The LCCSD scheme used here 
does not result in a  divergence, but
it is not perfect either. It generates only half of the screening of the 
long-range charge fluctuatons \cite{stohei}.

The weak-correlation expansion used here
looks very simple, but this is not at all the case
when
evaluated in a diagrammatic representation. In the linear equations
\ref{eq:expan0},\ref{eq:expan2}, the
interaction is included in both sets of terms. In a diagram representation, 
this means that infinite orders of diagrams are summed up. The 
LCCSD-approximation includes the Tamm-Dancoff approximation plus
all related exchange diagram corrections and also contains the Kanamori limit.
It does not yet contain the RPA-limit (plus all exchange corrections).
The RPA-limit is covered by the full CCSD-equations in eqs.\ref{eq:cc2},
\ref{eq:cc3}.

These findings also explain why 
reliable Green's function results for the
transition metals are rare. Only the
almost empty or filled band cases, i.e. the Kanamori limit \cite{58,59,60,61},
have been easily accessible.
For the case of Fe, one has been restricted to  MP2-calculations that
are more or less empirically renormalized \cite{un,lic}.
Only the DMFT has made a significant progress 
by the use of large scale Monte Carlo 
computations \cite{li2}.

     \section{Results for the non-magnetic Ground State}     
\label{sec:nm}
\subsection{Ground State Energies}

In the last section, some specific total energy contributions have been
analyzed. Here, we will discuss them in more detail.
The results discussed represent the bcc-case, and the ratio
${U \over W} = {1 \over 2}$ is used. Also $ \Delta J$ is disregarded
in the qualitative discussion for simplicity.

All total energies are related to an average interaction energy of
electrons with the same occupation with localized electrons and
without Hund's rule ordering:

\begin{equation}
E_0 (n_d)=(U-{2 \over 9}J)n_d(n_d-1).
\end{equation}
The Hund's rule energy gain for the ordered atoms is then
\begin{eqnarray}
E_{atom} (n_d)&=&-{7 \over 18}J \tilde{n}_d( \tilde{n}_d-1), \quad \mbox{where}
\ \tilde{n}_d             =  \left\{ \matrix{
                                     n_d \quad \mbox{for}\quad n_d \leq 5  \cr
                               (10-n_d)\quad \mbox{for} \quad n_d \geq 5  \cr}
.\right.
\end{eqnarray}

\begin{figure}[h,b,t]
\begin{center}

\includegraphics[width=6cm]{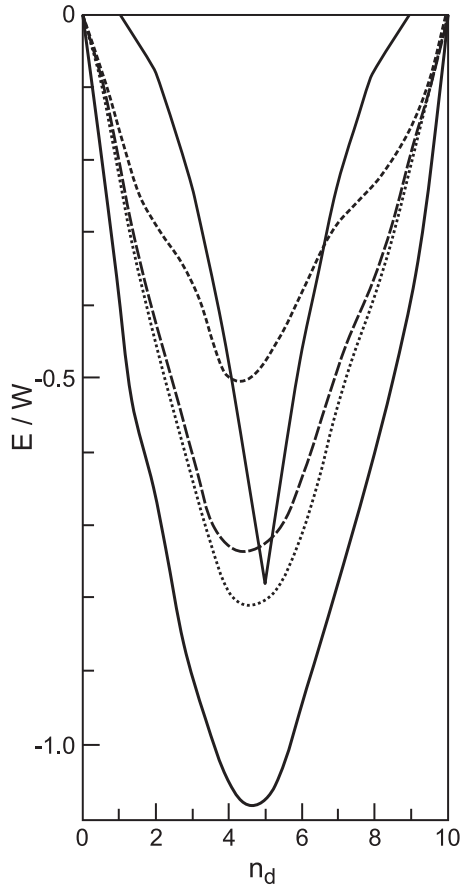} \\
\end{center}
\caption{Energies for various approximations of the ground state
in powers  of $W$ as a function of 
band filling $n_d$ for the bcc case. The upper solid line represents the
atomic energy, the lowest solid line the kinetic energy. The upper 
broken line represents the HF-energy, the lower broken line the result
with density correlations included, and the dotted line the final LA result.}
\label{fig2}
\end{figure}

The occupation dependence of this energy
is shown in Fig. \ref{fig2} (upper full line).
It is strongly peaked at $n_d=5$. 
It is compared
to the ground state energy for $H_0$ that is contained in the figure
as the lowest solid line. This figure indicates the binding due to 
the $3d$-electron delocalization. Disregarding slight shifts, the difference
between these two extremal curves is a good representation of the 
LDA binding energy
contributions of the $3d$-electrons. This can be seen
when comparing Fig. \ref{fig2} with the LDA binding energy
figure 1.1 in ref. \cite{18}. Two corrections should be made. The first one
is that the real atoms have  occupations differing from 
the solids ($Fe$-atom, $n_d=6$ versus $n_d=7.4$ in the model), and the
second one is that the bands are narrower for the heavier elements. By setting
$W=5eV$, a typical binding of $2.5eV$ arises.

The upper broken curve in Fig. \ref{fig2} represents the HF-energy. 
As already mentioned,
half of the band energy gain is lost. Even worse, the uncorrelated ground 
state is no longer binding in the occupation range from 4 to 7.
The situation is corrected when
the full correlation treatment is performed (dotted curve). The correlated
ground state is always bound, although only marginally at half-filling.
This is in rough agreement with experiment, where the d-orbital
contributions to the binding in $Mn$ are not larger than 1eV.
Actually, the difference between the non-interacting and the 
fully correlated result matches roughly the difference between the
LSDA and experimental binding energies for these cases (see again Fig. 1.1
in ref \cite{18}), and might well explain it, 
as will be discussed later.
The figure also contains the energy when spin correlations are omitted
(lower broken curve). As can be seen, the contributions of the spin 
correlations to the total energy are not large but nevertheless important.

\subsection{Correlation Functions}
\label{sec:cf}
\begin{figure}[h,b,t]
\begin{minipage}[b]{7.5cm}
{\centerline{\includegraphics{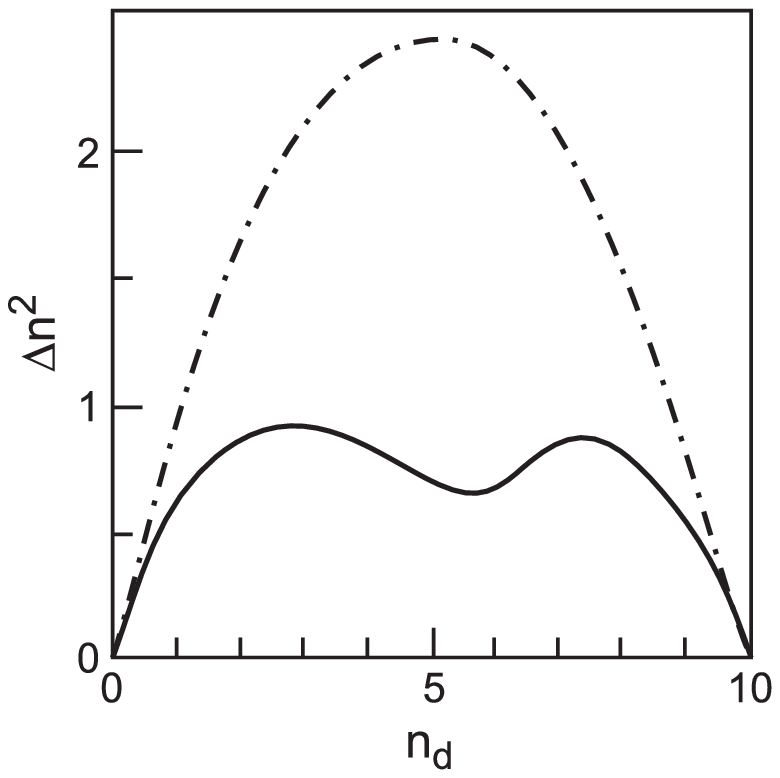}}}
\end{minipage} \hfill
\begin{minipage}[b]{7.5cm}
{\centerline{\includegraphics{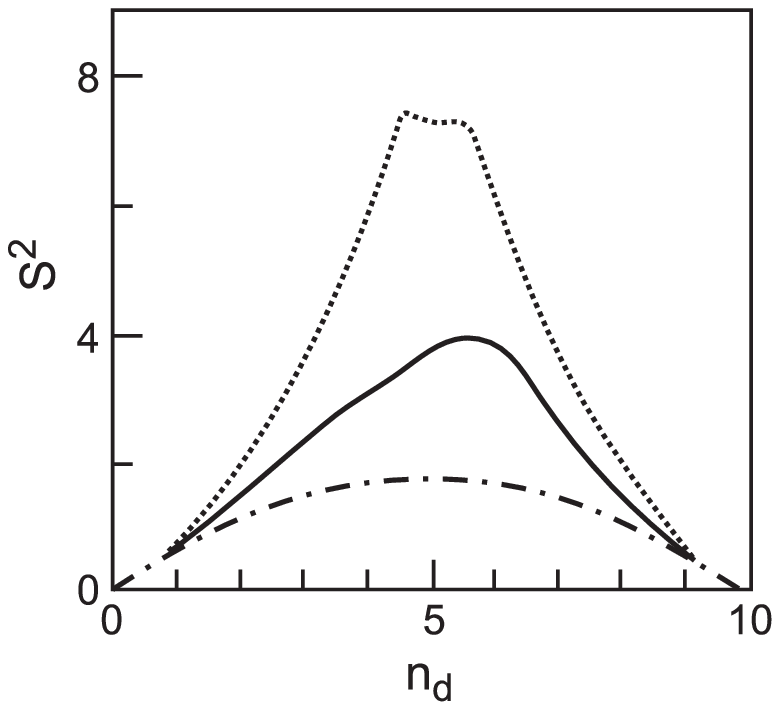}}}
\end{minipage}
\caption{Left figure:Charge fluctuations as function of d-band 
filling $n_d$ (bcc). Upper curve without correlations, lower curve with
correlations included \\
Right figure: Local spin correlations $S^2$ as function of d-band filling
$n_d$ (bcc). Upper curve: Atomic limit, lower curve:SCF-result. Full line:
LA-result (both from ref. \cite{st}).}
\label{fig4}
\end{figure}

From our calculations, local correlation functions 
for the transition metals were obtained for the first time. 
The effects of
the correlations are large, and should be basically experimentally 
accessible was not it for the yet lacking spatial resolution of x-rays, 
and for the too small energies of the neutrons. But it is still valuable
to discuss a few theoretical results.
The first correlation function is the atomic charge fluctuation $\Delta n^2$. 
The reduction of this quantity
due to correlations is shown in Fig.\ref{fig4}.
As can be seen, it is sizable, although the electrons are not strongly
correlated. This is due to the many available correlation channels.
Wenn turning to the spin correlation function $S^2$, one has to be aware
that the autocorrelation of the electrons leads to a finite value even
in the uncorrelated state. This is given as the lower broken line in the
right part of Fig. \ref{fig4}. Of importance is also  the fully localized limit
whose values of $S^2$ are given as the dotted line in this figure. The
correlation result is given as the full line. As can be seen, spin correlations
are strong but significantly smaller than in the localized limit. They
are almost halfways in-between, the relative change being 0.45, almost 
independent of band filling. This also indicates that the proximity of
the energy to the atomic limit at half-filling (see Fig. \ref{fig2})
does  not yet cause a resonance-like
correlation enhancement. 

This presence of relatively strong spin correlations poses
the question whether these can be treated in a good approximation
as quasi-local moments, whether they for example
already require a different timescale, or whether these correlations
decay as fast as the electrons move, and only form a polarization cloud around
the moving electrons. 

The answer cannot be directly obtained from ground state calculations.
There is, however, an indirect way to address this question, where one
includes short-range magnetic correlations into the
LA-calculation. There is no direct 
neighbor interaction in the Hamiltonian. Thus, a strong magnetic neighbor 
correlation would indicate the formation of local moments, at least
in cases when the non-magnetic ground state is only metastable, and
a ferromagnetic ground state exists. The calculation can be easily done
by adding neighbor spin operators to the correlation treatment.
This rules out a $R=0$-approximation. In the required computation, all
matrix elements $P_{ij}(0,l)$ with $l$ nearest neighbors are included, and 
all nearest neighbor
correlations are added up \cite{s85}. 

The discussion of the
obtained quantities requires some care because the non magnetic
ground state is a singlet. This implies that the positive magnetic
correlation function on the same site must be compensated by 
short-range antiferromagnetic correlations in order to obtain
$S_{tot}^2|\Psi_i\rangle =0 $, no matter whether the single particle (i=0)
or correlated (i=corr) ground state is concerned. 
Neighbor antiferromagnetic correlations 
have therefore no relevance as such, but only their eventual changes due 
to added degrees of freedom are of relevance. Consequently, we compare for
every filling the change in correlation with the maximal possible change,
namely the local moment formation. We discuss therefore the quantity
\begin{equation}
\Delta S^2_{\delta}={{C(\delta)-C_{0}(\delta)} \over {S^2_{loc}- S^2_0}}.
\end{equation}
Here, it holds that 
\begin{eqnarray}
C(\delta)&=&\langle \Psi_{corr} |S(l)S(l+\delta)|\Psi_{corr} \rangle \\ 
C_0(\delta)&=&\langle \Psi_0 |S(l)S(l+\delta)|\Psi_0 \rangle .
\end{eqnarray}
\begin{figure}[h,b,t]
{\centerline{\includegraphics[width=8cm,clip,angle=0]{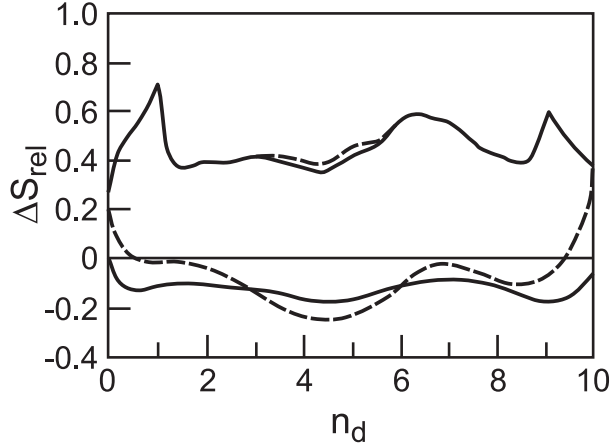}}}
\caption{Relative change of the spin correlation function as function
of band filling. Upper curves Change of on-site correlations, lower
curves change of neighbor correlations. Full lines without neighbor 
correlations, broken curves with neighbor correlations added (from ref. 
\cite{s85})}.
\label{fig5}
\end{figure}
Fig. \ref{fig5} displays these relative changes. In the upper part,
the relative change of the on site spin correlation is given, in the
lower part the one of the neighbor functions. The solid line displays
the result without neighbor correlations that had just been discussed.
The upper curve indicates on the average 45 percent of the maximal 
correlation, and the lower curve indicates the required antiferromagnetic 
re-ordering due to the on-site correlations in the neighbor function.
This calculation was performed for the bcc case. As can be seen, the ratio
is a little smaller than ${1 \over 4}$. Therefore, additional longer range 
compensation effects are expected. 
The most interesting result is the changes due to 
added neighbor correlations. The resulting curves are 
given in Fig. \ref{fig5} as dotted lines. 
These changes are very small. However, it is interesting that 
they recover the expected trends correctly. They indicate
a tendency towards antiferromagnetism only around 
half-filling (from occupations
of $3.5<n_d<6.5$). Apparently, the magnetic susceptibilities are slightly 
enhanced for the proper magnetic ordering. However, the moments themselves
do not change at all, except around half-filling. Here,
the stability of the non-magnetic state is smallest,
as discussed above. It should be noted that for this choice of parameters, 
the  stable ground state
is ferromagnetic for band fillings of $n_d>7.0$. 

To conclude, on-site correlations on neighbor atoms
do not support each other. Barely noticable neighbor correlations form.
These results strongly contradict a local moment assumption.
The energy gain due to the added operators is small, it
amounts to less than 100K per atom for Fe.
Consequently, the strong on-site correlations cannot be interpreted
as quasi-static local moments. 
Magnetic ordering restricted to nearest neighbors
does not exist in these compounds. If magnetic order
exists - and it must exist - then only for domains considerably larger
than a single atom and its neighbors. This demonstrates again that the
electrons in the transition metals are delocalized, and that spin fluctuations
can only exist for small moments $q$, as experiments demonstrate
(the stiffness constant of $Fe$, for example allowes magnetic excitations 
with energies smaller than $T_c$ only for moments 
$q$ smaller than one fifth of the Brioullin zone).

If there is really a need to 
adress i.e.
the strong magnetic scattering above $T_c$ for $Ni$ \cite{52} by methods 
extending beyond a critical Stoner-enhancement that was computed
in ref. \cite{cstei},
then this can only be done by long-range spin or 
order parameter fluctuation theories \cite{54,55,53}.

\subsection{Compton Scattering}  
 
While experiments have not yet been able to provide informations about 
correlation
functions, they have succeeded for another quantity that displays correlation
effects: the density distribution in momentum space, $n({\bf k})$.
The scattering intensity $I({\bf q})$ measured in Compton-scattering
is given by the integral over
all densities $n( \bf k)$ with ${\bf kq=q^2}$. The variation in
$I({\bf q})$ with direction ${\bf q}$ provides a direct measure of the
anisotropy of the Fermi surface \cite{c1}. These experimental results 
are in good qualitative agreement with Fermi surfaces obtained in LSDA
for $Cu$ \cite{c1}, $Va$\cite{c2,c3}, $Cr$\cite{c3}, and $Ni$\cite{c4} 
with exception
of a constant scaling factor. For $Fe$\cite{c5,c6}, the agreement is less
good.

\begin{figure}[h,b,t]
\begin{minipage}[b]{7.5cm}
{\centerline{\includegraphics{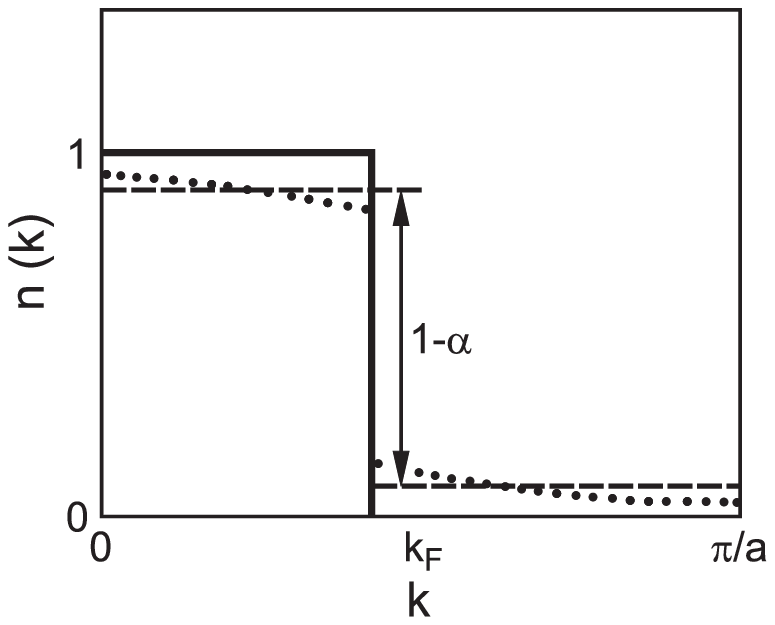}}}
\end{minipage} \hfill
\begin{minipage}[b]{7.5cm}
{\centerline{\includegraphics{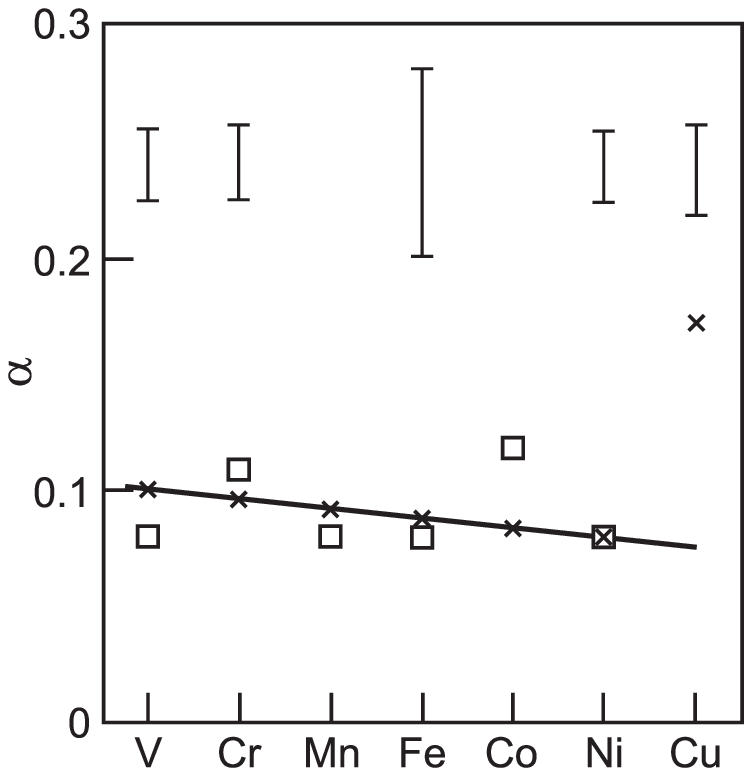}}}
\end{minipage}
\caption{Left figure:Qualitative picture of changes in occupation around the
Fermi energy due to correlations. Dots represent the real behaviour, and
the dotted line the average change.\\
Right figure: Values of the quantity $\alpha$ in the left figure for
different transition metals. The bars represent the values deduced from 
experiment in comparison to LSDA-results. Crosses give an homogeneous
electron gas estimate $a_{hom}$, and squares give the contributions
$\alpha_{LA}$ due to the atomic correlations of the screened d-electrons
alone to this value taken from \cite{s952}.}
\label{fig6}
\end{figure}

This constant scaling factor provides a measure of the correlation correction.
In the single-particle approximation, all states with energies smaller
than the Fermi energy are filled and the others are empty.
This implies a maximal step at $n({\bf k_F})$. This result 
is changed by correlations.
The changes are qualitatively depicted in the left part of Fig. \ref{fig6}.
For simplicity let us assume
that
correlations cause a constant shift $\alpha$ in occupations
for the occupied and unoccupied parts of the partially filled bands. Then
this correction $\alpha$ can be directly extracted from
the scaling factor by which 
experiments 
(correlations included) and single-particle calculations
(no correlations included) differ.

Fig. \ref{fig6} (right part) contains the
values for $\alpha$ 
for the mentioned transition metals and for $Cu$
as deduced from experiment. As can be seen this
reduction amounts to 20-25 $\%$ in all cases.

This reduction $\alpha$ can be compared to its counterpart
$\alpha_{hom}$ obtained for a homogeneous
system with the same average density.
For its derivation, we refer to \cite{c1,s952}.
When taking for $Cu$ a single valence electron only
(i. e. considering the d-electrons as part of the core), then the experimental
value is regained. This indicates that as far as this property is 
concerned,
the 4s (and 4p) electrons correlate as a homogeneous system
of the same average density.
 For the transition metals, on the other hand,
the d-electrons need to be incorporated
into the estimate. The resulting $r_s$ value is much smaller
than in the case of $Cu$ which implies - within the theory of the 
homogeneous electron gas - a smaller reduction of the occupation.
The latter  amounts to
less than half of the correlation effects determined experimentally.

Within our scheme,
the reduction $\alpha_{LA}$ is obtained from the change
of the expectation value of $H_0$, the so-called
kinetic or band energy,  with correlations. For the uncorrelated ground
state, it holds that
\begin{equation}
E(band)_0=\langle\Psi_o|H_0|\Psi_0\rangle=\sum_{\nu \sigma}
\int_{e_{\nu}({\bf k})\leq e_F(\sigma)}d^3 {\bf k}
e_{\nu}({\bf k})
\end{equation}
while the band energy of the correlated ground state is
\begin{equation}
E(band)=\langle\Psi_{corr}|H_0|\Psi_{corr}\rangle=(1-\alpha_{LA})E(band)_0
\label{compt5}
\end{equation}
because the  model Hamiltonian is constructed such that $Tr(H_0)=0$.
Thus, the quantity $\alpha_{LA}$ also represents the relative change of the
band or kinetic energy of the model by correlations.
For the magnetic cases, the values for the magnetic ground state are
selected. The interaction parameters are chosen so that they correspond
to the actual transition metals. The values found indicate again that
the electrons are weakly correlated. Only 10 percent of the kinetic
energy is lost due to atomic correlations.

The restriction to this model implies that
 only a part of the correlation corrections $\alpha$
can be obtained, i.e. the one 
which arises from
the atomic correlations due to  the strongly screened
atomic d-electron interactions.
The 
contributions of the screening itself to 
the momentum density, for example, are
not included in this estimate.
 As can be seen from Fig. \ref{fig6},
the particular atomic correlation contributions
$\alpha_{LA}$ alone as derived from the model computations
are as large as the total homogeneous electron gas values $\alpha_{hom}$.
Therefore, they must to a large extent be 
neglected in a homogeneous electron
gas approximation as was explained before.
They amount to almost half the experimental 
value $\alpha$ and can therefore explain
the largest part of the deficit of a homogeneous electron gas treatment. 

Correlations are included in  a homogeneous electron gas approximation
when LSDA calculations are performed.
Such calculations must therefore lack a satisfactory description
of these atomic correlations
for the case of the transition metals.

     \section{Results for the magnetic Ground State}     
\label{sec:mag}

\subsection{Parametrization of the Hubbard Hamiltonian}

Of main interest in the case of the magnetic transition metals is the magnetic
moment itself. So far, for Hubbard models the
interaction parameters were always chosen such that the experimental
magnetic moment was obtained. It is an indication
for the accuracy of the different discussed correlation treatments that
these parameters are now closely related. Global differences
for the most typical value $U+2J$ are not larger than 10 percent, and
are connected with band structure differences between 5-band and 9-band
models. This holds true as long as the interaction is restricted.
When also interactions on and between the $4s,4p$-orbitals are included,
then the $3d$-interaction also needs to be enhanced and the screening 
of the latter interaction by the $4s,4p$-electrons is explicitly
covered. 

The values for $W$ and $U$ for our model are given in table \ref{table1}
together with the moment used.
\begin{table}[h]
{
\begin{center}
{   \begin{tabular}{|l|rrr|}
 \hline                                                                        
&$Fe$&$Co$&$Ni$\quad\\
\hline
$n$&7.4&8.4&9.4\\
$m$ &2.1&1.6&0.6\\
$W$&5.4&4.8&4.3\\
$U$&2.4&3.1&3.3\\ \hline
\hline
\end{tabular}}                                                                 
\end{center}}
\caption{\protect Parameters for our model (energies in eV)
(from ref.\cite{os}).}
 \label{table1}
\end{table}                                                        
In all cases, the ratio $J=0.2*U$ was kept. Only for the case of $Fe$ was
$U$ unambiguously determined from the magnetic moment. In the other cases
we had only lower limits which were 2.6 and 3.1eV for $Co$ and $Ni$, 
respectively. Below, we will present a comparison between the values of $U$
obtained here, and the ones obtained from other sources.

\subsection{Dependence of Magnetism on Degeneracy}

The degeneracy of the energy bands of the transition metals is
of vital importance for magnetism itself, and also imposes strict
boundary conditions on the possible treatments. To explain this, 
the magnetic energy gain and its magnetic moment dependence are analysed 
as a function of the represented method $i$ for the actual moment $m_0$.
\begin{equation}
\Delta E_i(m_0)=E_i(m_0)-E_i(0).
\end{equation}
This energy gain as a
function of magnetization is rewritten in the following form
\begin{equation}
\Delta E_i(m_0)={1 \over 4} \int_0^{m_0^2}D(m)dm^2 -{1\over 4}
\int_0^{m_0^2}I_i(m)dm^2. \label{sto1}
\end{equation}
Here, the first term describes the loss in (non-interacting) band or
kinetic energy.
It holds that 
\begin{equation}
D(0)={1 \over n(E_F)}
\end{equation}
is the inverse total density of states per spin at the Fermi energy. 
Its generalization
for finite m is simple and can be found in ref. \cite{soh}.
The second part describes the interaction energy gain and is defined
by this function. It is a generalized Stoner parameter.
The optimal magnetic moment in approximation $i$ is defined by the condition
\begin{equation}
D(m_0)=I_i(m_0).
\end{equation}
For $m_0=0$, this is the standard Stoner criterion, and the
Stoner parameter $I_i$ is the limiting $I_i(O)$.
For a system with orbital degeneracy $N$ and a 
Hubbard interaction
in the form of eq. \ref{eq12}, it holds in HF-approximation
\begin{equation}
I_{SCF}(M)={1 \over N }(U+J(N+1)).
\end{equation}
If we assume a structureless density of states with a bandwidth
$W$, then $D(0)={W \over N}$, and for the single band
model the Stoner criterion in SCF-approximation reads
$U+2J=W$. In a single-band Hamiltonian, the interaction terms are
usually condensed into a single $U$, but for the degeneracy treatment
we will stay in our notation. Ferromagnetism in a single-band
system can therefore only be expected for a strong interaction
with $U \geq W$ where correlations are important. 
Correlations however strongly diminish
$I$ from $I_{SCF}$, and shift the onset of magnetism to an even larger 
interaction. Thus, if spurious magnetism due to 
peaks in the density of states is disregarded,
itinerant magnetism with large moments and weak correlations can arise
only for highly degenerate systems. 
This is why we could obtain magnetism for 5-band systems with rather weak
interactions of ${U \over W}\geq 0.5$. 
The atomic exchange interaction
$J$ is the relevant quantity in this respect, and it requires an adequate
treatment. 
Note that for the case with the 
smallest interaction, $Fe$, magnetism is strongly supported by a
peak in the density of states, and that $Fe$ does not become fully magnetic. 

\begin{figure}[h,b,t]
{\centerline{\includegraphics[width=8cm,clip,angle=0]{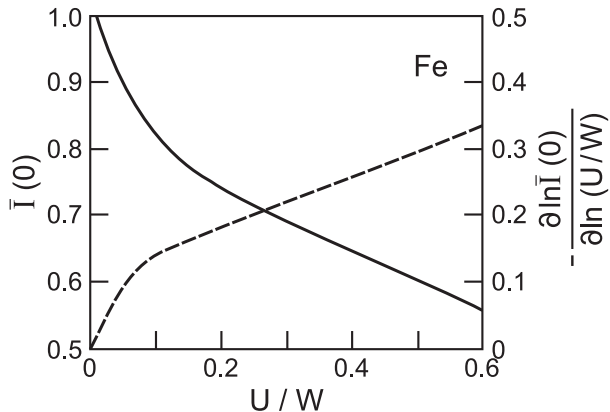}}}
\caption{$I(0)$ as depending on the ratio ${U \over W}$ for Fe.
In addition, the logarithmic deviation is given (from ref.\cite{soh}).}
\label{fig8}
\end{figure}

We had already mentioned that the treatment of degenerate band systems
puts strong additional demands on the many-body methods used. This
also is the case for magnetic properties. Fig. \ref{fig8} contains, for
the bcc $Fe$-case, the correlated Stoner Parameter I(0) as a function
of interaction $U$, renormalized by the SCF-Stoner parameter
(roughly $(U+6J)$) to $\bar I$. For $U$=0, $\bar I(0)$ is therefore equal to 1,
and it decreases due to correlation corrections.
As can be seen, there is first a fast decrease, but then, at ${U \over W}=0.1$,
a rigorous slowing down of the screening occurs. From the logarithmic 
derivative
it can be seen that the exponent changes from 
2 to ${1 \over 2}$. This is the point where a second-order
perturbation treatment is no longer sufficient, and a better approach is 
necessary.

Due to these experiences we decided to include in this review neither
contributions
that used false degeneracies nor contributions that
are unable to treat degenerate systems well. Therefore, we did not
cover
work that is based on MP2 or lowest order
diagram techniques. Besides the LA only the full 
$CI$-method given in refs \cite{web,web1}, Kanamori t-matrix
applications,
and the first full applicaton
of the DMFT to the transition metals \cite{li2} remained.

\subsection{Magnetic Energy Gains, Stoner Parameter and $T_c$}

Besides the moment that was taken from experiment, 
the most basic ground state quantity connected with the broken
symmetry 
is the 
magnetization energy gain $\Delta E$. 
\begin{table}[h]
{
\begin{center}
{   \begin{tabular}{|l|rrr|}
 \hline                                                                        
&$Fe$&$Co$&$Ni$\quad\\
\hline
$\Delta E_{HF} [eV]$&0.56&0.43&0.12\\
$\Delta E_{corr}$&0.15&0.13&0.03\\
$\Delta E_{LSDA}$&0.28&0.10&0.08\\
\hline
$T_c(LSDA)[K]$&4400&3300&2900\\
$T_c(DMFT)$&2000&&700\\
$T_c(exp)$&1040&1400&631\\ 
\hline
\end{tabular}}                                                                 
\end{center}}
\caption{\protect Magnetic
energy gain in the HF-approximation and for the correlated 
ground state \cite{os}
and from  LSDA calculations \cite{gunn1}. Stoner-transition temperatures
from LSDA calculations \cite{gunn1} and DMFT \cite{li2} and 
the experimental transition temperatures.}
 \label{table2}
\end{table}                                                        

The  values of the energy gain are given in table \ref{table2} for our 
calculations in the
single-particle approximation, for the full treatment of correlations,
and for LSDA-calcluations \cite{gunn1}.  A comparison of $\Delta E_{HF}$
and $\Delta E_{corr}$ shows how correlations reduce the magnetization
energy gains. Be aware that the interactions employed are already 
strongly reduced
screened interactions. 
When comparing the model and LSDA magnetic energy gains,
then the LSDA quantities come out twice as large for $Fe$ and $Ni$.
Without caring for any specific dependences on details of the density of
states, this would imply a corresponding reduction in the Stoner $T_c$.
An exception is the case of $Co$ but here we had possibly chosen
a model interaction that was too large for reasons explained below.

The differences between the LA and the LSDA results can be understood
by a discussion of the Stoner parameter $I(m)$. In the SCF-approximation,
it holds for the degenerate band case
\begin{equation}
I_{SCF}(m)={1 \over 5}(U+6J).
\end{equation}
(the terms $\Delta J$ are disregarded). $I_{SCF}$ is 
independent of magnetization.
The function $\Delta E_{LSDA}$
contains the same expression for the kinetic energy
as the quantities defined above because in this approximation the
uncorrelated kinetic energy of the reference wave function is used,
and also  $I_{LSDA}$ turns out to be
independent of magnetic moment. Even more important,$I_{LSDA}$ is 
also essentially
independent of the kind of transition metal atom and its environment.
It holds (within 10 percent variation) $I_{LSDA}=0.9eV$ \cite{gunn1}.

While in LSDA the uncorrelated kinetic energy is used
and the losses in band energy due to magnetism are large, correlations reduce 
these losses for $Fe$ by 30 percent \cite{soh}. These corrections are 
included in $I(m)$. In particular, if
$D(m)$ changes with $m$, then a partial compensating change must occur in this
function. This holds true for $Fe$, where it is well-known that
$D(m)$ strongly rises with $m$ and cuts off the magnetic moment
before the maximum. One finds that $D(m_0)= 1.6*D(0)$ \cite{soh}.
Correspondingly,
$I_{corr}(m)$ must contain a correcting change. Since 
$I_{corr}(m_0)=D(m_0)$, it holds that $I_{corr}(m_0)=1.2I_{corr}(0)$
due to this effect alone.

\begin{figure}[h,b,t]
{\centerline{\includegraphics[width=8cm,clip,angle=0]{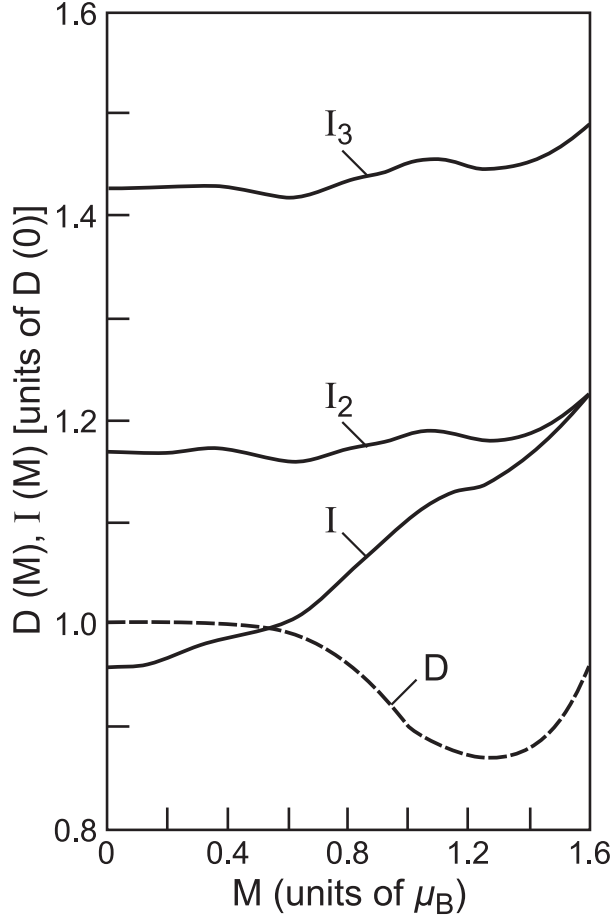}}}
\caption{Kinetic energy loss function $D(M)$ in comparison to the Stoner 
parameter $I(M)$, obtained correctly (I), without inclusions of spin 
correlations $I_2$, and without any correlation corrections for 
the $J$-contributions, $I_3$ (from ref. \cite{soh}).}
\label{fig7}
\end{figure}

There is a further interesting  correction that
can be seen immediately  for $Co$ where the density of state is constant and
causes no $m$-dependences in $I(m)$.
Fig. \ref{fig7} displays the 
function $I_{corr}(m)$ (called $I(m)$ in the figure) in comparison to $D(m)$.
As can be seen, it increases sizably with $m$. The origin is due to
spin correlations. When these are turned off in the correlation calculation,
the resulting quantity $I_2(m)$ no longer displays $m$-dependencies. Spin
correlations are most relevant in the non-magnetic state and die out
in the full magnetic limit. Also contained in the figure is the
quantity $I_3(m)$ which is obtained when all interaction contributions
originating from $J$ are treated in the $HF$-approximation.

This proper treatment of Hund's rule effects, 
which corrects the Stoner parameter,
strongly contrasts to fictitious disordered local moments that
result from a locally symmetry broken HF-treatment of the exchange 
interaction. 
If such local moments existed in reality,
then they would from the beginning prefer
to order and to enhance the kinetic energy of the electrons.

The effect seen here is a quasi 'anti-disordered' local moment effect.
It definitely squeezes $Co$ into a first order magnetic phase transition. The
non-magnetic ground state for $Co$ is even metastable at $T=0$. It
requires a polarization beyond a critical size to destabilize it towards the
ground state. Consequently, we would expect that the LSDA value for
the transition temperature is not only overestimated due to a missing
correlation correction of the kinetic energy loss, but that the
true first order transition is considerably below that corrected quantity.
We have not performed thermodynamic estimates but it looks as if
the experimental $T_c$ for $Co$ might be reached this way.
Our original choice of $U$ for $Co$ \cite{os} had been motivated 
by the wish to bring $Co$ close to a second order phase transition.

Where $Ni$ is concerned, our results indicate a strong reduction of
the magnetic energy in comparison to LSDA. For this case, a proper 
thermodynamic treatment has been made in DMFT \cite{li2}. 
The Hamiltonian was very
close to the one that we used, and here also a local mean field 
or Stoner-theory was applied. The outcome was a $T_c$ slightly
above the experimental value. Also the critical spin-fluctuations
due to Stoner-enhancement just above $T_c$ were in good qualitative 
agreement with experiment. Corrections by spin-wave fluctuations
that were disregarded in the DMFT calculations
are definitely of relevance but have apparently little effect. 
One should remember, though, that the values of the interaction $U$ are not
unambiguously fixed for the $Ni$-model calculations. The agreement
of the DMFT calculations and experiment might originate in part from
a value for $U$ that is slightly too small. 
On the other hand, there is again the
trend that $I(m_0)$ is 20 percent larger than $I(0)$\cite{soh}, 
in part due to the
spin correlations, and in part due to a small reduction of the change
in $D(m)$. Since these two corrections are missing in LSDA-calculations, one
would expect that the
true Stoner-$T_c$ is considerably smaller than the LSDA-$T_c$.

For $Fe$, relative changes in the partial $n_{e_g}$ and $n_{t_{2g}}$
occupations with magnetization arise as artefacts of the five-band
model. These cause an enhancement of  $I_{SCF}(m)$ for small $m$.
Being only partly screened by correlations, this partially compensates
the corrections for small $m$ mentioned above.
Nevertheless, a reduction of the magnetic energy by half 
was still found \cite{soh}.
However, this should, due to the pathology of the $Fe$-density of
states, only lead to a reduction of $T_c$ by less than half \cite{gunn1}. The 
DMFT-calculation for a nine-band
Hamiltonian for Fe produced in fact a $T_c$ of 2000K \cite{li2} , which is
unsatisfactory. Here, the interactions were unambiguously fixed,
and the deficiencies of the five-band model were absent.

Thus $Fe$ is the only case that still poses a major problem.
There are two  possible scenarios for its solution:
either the phase transition
in $Fe$ is not at all correctly described by a Stoner-like picture but
rather by spin-wave fluctuation theories (for the earliest ones, see
refs. \cite{54,55,53}), or, alternatively, the starting point, the 
rigid band system derived from LSDA together with local 
Hubbard-interactions is insufficient. Very recently, we obtained
evidence \cite{sun} that
the latter case holds true. Its discussion is necessary for an
understanding of the limits of a Hubbard model treatment 
but goes beyond the scope of this contribution.
It will therefore be shortly addressed  at the end.  

\subsection{Anisotropic Exchange Splitting for Ni}

For ferromagnetic $Ni$, an interesting anisotropy occurs. The
majority bands are completely filled. The charge in the minority
bands is not equally distributed among the different $3d$-orbitals,
but charge is missing almost exclusively from the $t_{2g}$-orbitals
which form the most antibonding band states. This anisotropic 
charge distribution is, to a smaller amount, already present for
the non-magnetic state without interaction
(occupations of 0.98 or 0.92 for the $e_g$ and $t_{2g}$ orbitals,
respectively). In the SCF-approximation, already for the non-magnetic
state, an anisotropic crystal field exists. Even more interesting,
for the ferromagnetic state, an anisotropic exchange splitting builds
up. With exchange splitting, we refer to the difference in majority and
minority crystal field terms. In the SCF-approximation, it holds
for the splitting $\Delta_{SCF}(i)$ (with terms in the Hamiltonian 
$\Delta J$ disregarded)
\begin{equation}
\Delta_{SCF}(i)=(U+J)(n_{i\uparrow}-n_{i\downarrow})+Js_z
\end{equation}
where
\begin{equation}
s_z=\sum_i(n_{i\uparrow}-n_{i\downarrow}) \ \ .
\end{equation}
While the second term is isotropic and amounts to 0.4eV for $Ni$,
the first one contributes a splitting of 0.7eV only for the $t_{2g}$-orbitals.

With correlations, the interaction effects are partially screened. 
Also, a single-particle potential is no longer unambiguously defined.
We obtained approximate values from our ground state calculations
by keeping the  correlation operators restricted to two-particle excitations,
and by energy optimizing
correlated states starting from different single determinant trial states 
that were each generated with particluar exchange splittings. 
The optimal trial state
determined the ground state exchange splitting.
We obtained for $Ni$ exchange splittings of $\Delta_{corr}(e_g)=0.15eV$ and
$\Delta_{corr}(t_{2g})=0.57eV$\cite{os}. Actually, here interaction 
terms $\Delta J$ contributed. Without them, the splittings
would be 0.27 and 0.50eV, respectively. These changes indicate 
a screening of the $J$ contributions by almost half, and of the 
$U$-contributions by more than half. Exchange splittings from LSDA
come out isotropic and amount to 0.6eV.

These anisotropies were measured in angle resolved photoemission
experiments \cite{him1,plu} and came out as 0.1 or 0.4 eV, respectively.
The agreement with our values is good; in particular if one
takes into account that we computed our splitting for non-renormalized
bands while in photoemission experiments the 15-20 percent 
mass renormalization due to many-body effects is included \cite{him,cooke}.

Many-body calculations for quasiparticle properties had actually been
done before for a five-band Hamiltonian with very similar interactions
(lacking the $\Delta J$ contributions). This calculation was
based on the Kanamori t-matrix approximation, and had obtained splittings
of 0.21 and 0.37 eV, with which our results agreed well. That computation
had not only obtained the anisotropic splitting but also reasonable
band renomalizations of 15 percent and even the experimentally seen
shake up peak \cite{58,59}. Very recently, ground state calculations
for a Hamiltonian similar to ours were performed, 
this time again within the $R=0$-approximation
but for a 9-band model Hamiltonian, and by a full CI-calculation. The agreement
for the exchange splitting was again very good (0.16 and 0.38 eV), but
this time also relativistic contributions were included, and the experimental
Fermi surface was reproduced with high quality \cite{web1}. Here, the 
anisotropic exchange splitting plays a big role, 
and the Fermi-surface of the LSDA is false.

In the computations,
such anisotropies in the exchange splitting did not show up for $Co$, but
we found them for $Fe$, too. In Fe, the $e_g$ orbitals carry a larger
moment because the majority states are completely filled, but the
minority bands are less populated than the $t_{2g}$ analogs. We had found
splittings of $\Delta(e_g)=1.74eV$ and $\Delta(t_{2g})=1.30eV$. 
LSDA-calculations obtain an isotropic  splitting of 1.55eV, and experiments
obtain 1.45eV\cite{him} but cannot resolve an anisotropy.

   \section{Interaction Parameters U}

\begin{figure}[h,b,t]
{\centerline{\includegraphics[width=8cm,clip,angle=0]{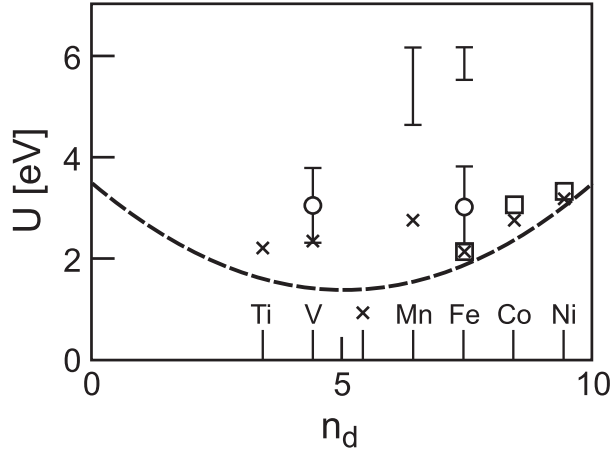}}}
\caption{Values of the interaction $U$, obtained for different transition
metals as a function of band filling. The values obtained by the LA are
given as squares, the ones obtained from core spectroscopy \cite{saw}
are given by crosses. The full curve indicates the occupation dependence
when a residual neighbor interaction $V$ is included. Bars with circles
indicate estimates of $U$, obtained from ab-initio 
LA calculations\cite{sun}, bars without circles estimates from frozen 
charge approximation computations\cite{cgunn,gudra}}
\label{fig9}
\end{figure}

It is of interest to compare the interaction parameters $U$ obtained
for $Fe$ to $Ni$ to parameters obtained by other means, and to try to
gain a more global understanding. Fig. \ref{fig9} contains these 
parameters as functions of the band filling.

Spectroscopy experiments for transition metal impurities in noble
metals
were interpreted with the help of atomic interactions \cite{saw}. 
The values of
the resulting quantity, $U$ (or $F_0$, as it is called in ref
\cite{saw}), are given in  Fig \ref{fig9}, too.
They are in very good agreement with the $U$ obtained
by the LA, in particular if one
considers the different environments. This indicates that the screening
patterns must be similar, and must also originate from the $4s,4p$-orbitals
on the one hand, or from the $5s,5p$-orbitals on the other.
In the impurity case, results exist for a whole range of transition
metal impurities.
The maximal reduction is obtained for the half-filled case (Cr)
with $U=0.9$eV taken from spectroscopy results, while
for less then half filling, the interaction increases again \cite{saw}.

The occupation dependence of $U$ can be understood with the 
help of
a residual neighbor electron interaction $V$.

It has been worked out before for a single-band model \cite{s951}, how
a Hamiltonian with on-site and neighbor interactions,
\begin{equation}
H_{int}=U_0\sum_i n_{\uparrow}(i)n_{\downarrow}(i)+{V \over 2}\sum_{(i,j)}
a^{\dagger}_{\sigma}(i)a^{\dagger}_{\sigma'}(j)
a_{\sigma'}(j)a_{\sigma}(i),
\end{equation}
can be mapped into the Hubbard interaction
$H_{int}=U \sum_in_{\uparrow}(i)n_{\downarrow}(i)$.
The sum
over$(i,j)$runs over the $z$ nearest neighbors $j$ for each atom $i$.
Here, it is implicitly assumed that all
longer-range interactions are equal to zero.
For the Hamiltonian with only on-site interactions, the only 
meaningfull response to the interaction can be condensed into the
expectation value $\langle n_{\uparrow}(i)n_{\downarrow}(i)H_{int}\rangle$.
It is therefore plausible to choose $U$ so that it yields the
same response as the pair $U_0,V$.
Assuming for $\Psi_0$ the following relations
for the single particle density matrix, 
$P(0) =<|n_{\sigma}(0)|> =n$, $P(0,l) =
<|a^{\dagger}_{\sigma}(0)a_{\sigma}(l)|> \simeq p$ for ($l$)nearest neighbors,
and zero elsewhere, one obtains
\begin{equation}
<|\overline{n_{i\uparrow}n_{i\downarrow}}H_{int}|>=
U_0((n(1-n))^2+zp^4)-2Vzn(1-n)p^2  \ \ ,
\label{modh0}
\end{equation}
and
\begin{equation}
U=
U_0
-{{2Vzn(1-n)p^2} \over{
(n(1-n))^2+zp^4}}=U_0-\alpha V
  \ \ .
\label{modh1}
\end{equation}
The parameter $U$
depends therefore on band filling as well as 
on the number of nearest neighbors. 
For almost empty bands, it holds that $p \simeq n$ since
 $k_F\ll{1\over{|\vec{R}_i-\vec{R}_j|}}$
and consequently $U\simeq U_0$.
At half-filling, it follows from $P^2 = P$, that
$n \simeq n^2 + zp^2$ and therefore
$\alpha=2z/(z+1)$.
For almost completely
filled
bands
it holds that $\alpha \sim \delta n$, where $\delta n$ represents the
number of holes.
The curve drawn into Fig \ref{fig9} is based on an $U_0=3.5eV$ and
a $V=1.2eV$. It is deliberately put 0.5-1.0eV below the data points.
Apparently, a pair of interactions $U=4.2eV, V=1.2eV$ can replace the
so far independent terms for the specific elements. 

Accepting such a residual interaction $V$
for the transition metals resolves a further problem.
The LDA band width for transition metals is known to roughly
agree with experiment for $Fe$ but to be 
 too large for $Co$ and $Ni$ \cite{cooke,him}.
The latter deficiency has been understood to arise from
correlations caused by the on-site interactions \cite{58}.
From model calculations with the LA \cite{os,soh}, 
it can be deduced that the reduction
of the band width $\delta W$ due to $U$ is similar in all three (!) cases.
The apparent discrepancy for the case of $Fe$ can be
resolved
when a neighbor interaction $V=1.2$eV is included.
The exchange broadening $\delta W$
due
to $V$ 
increases
the band width of Fe by 
10\% which partly compensates 
the correlation correction due to $U$.
The exchange corrections for $Co$ and $Ni$ are smaller since it holds again
that
$\delta W \sim V \delta n$.

 Effective local interactions $U$ 
were computed from LDA frozen charge calculations for
the transition metal $Fe$ and for $Mn$ impurities in $Ag$
\cite{cgunn,gudra}. They are plotted in Fig. \ref{fig9} as
well. Apparently, these
results do not depend
on band filling. They are of the same size as 
for the case of $Cu$ with a completely filled 3d band \cite{cgunn}.
There, they match experiment.
These frozen charge calculations actually computed the interaction
costs
to bring two electrons together from infinite distance but not
the change in interaction from a neighbor site to the same site
which is the relevant quantity
in a half-filled band system. It is thus
not astonishing that
frozen charge LDA calculations are
unable to treat band filling effects 
on effective local interactions.

The convolution of a longer range interaction into an on-site term
alone
is not sufficient to determine a model $U$. A model 
always lacks degrees of freedom that are present in the ab-initio
calculation. If degrees of freedom were removed, then there is an
alternative procedure to obtain
a residual local interaction $U$. It 
is to require that particular correlation properties of
the model are identical to the same ab-initio quantities. 
Due to the restriction of the model interaction to atomic terms,
the relevant properties are atomic correlations.
For the cases discussed here, the proper representative is
the change of the atomic charge fluctuations $\Delta n^2$. For
the model, this quantity was discussed in section \ref{sec:cf}.
Exactly the same quantity can be calculated from ab-initio calculations.
There, atomic orbitals are unambiguously defined, the same operators
are included into the correlation calculations, and the same correlation
function is available. $U$ is then chosen so that the model correlation
function matches the ab-initio result. First applications have been
presented in ref. \cite{s951}. From first ab-initio calculations,
correlation functions for $V$ and $Fe$ are available \cite{sun}. 
They lead to
values of $U$ that are also included in Fig. \ref{fig9}. The big error
bars originate from a mismatch of the five-band Hamiltonian to the
correct atomic orbitals. We did not want to enter poorly described
$3d$-orbitals with $4s$- and $4p$-tails into the ab-initio calculations.
On the other hand, lacking a general tight-binding program,
we could not improve the model. Also, in our ab-initio calculation,
we did not treat the very short-range part of the correlation hole well.
Thus, our ab-initio results rather represent an upper limit to the $U$.

Even for these error bars, the results show that an
unambiguous  convolution and
condensation of the full Coulomb interaction into meaningful model
interactions is possible. So far, model interactions were
always chosen to fit a model to experiment. We had done so
for the transition metals, and had obtained good 
agreement, but we had encountered other cases where a particular
physical effect was incorrectly connected solely to an on-site interaction,
and the resulting fit led to a wrong $U$. The case we have in
mind is polyacetylene whose bond alternations also depends on
interactions but not soleley on a local term $U$ \cite{ks,s951}. 

Of interest is also the case of the high $T_c$-superconducting 
compounds where we had
performed such an analysis \cite{s98,st02}.

\section{Relation of the Hubbard Model Results 
to the LSDA}

In the previous chapters, we had shown how difficult an
adequate treatment even of model interactions is. This makes it
even more astonishing that LSDA-calculations managed
to get sizable parts of magnetism correct, in particular the
magnetic moments of the transition metals. To a certain degree
this results from the fact that the $3d$-electrons are not too
strongly correlated, and that for the delocalized electrons 
extended-H\"uckel features prevail, which the LSDA describes well.
This may explain the cases of $Co$ and $Ni$, where the
magnetic moment is maximal, provided the $3d$-band occupation is correct.

But as our evaluations have made clear,
the correct magnitude of the magnetic moment of $Fe$ required
an accuracy that the LSDA cannot possess. Consequently, the correct
result can only arise due to a chance compensation of a set
of errors.
As was shown, the magnetic moment is determined by the Stoner-Parameter I,
which depends equally on the local interactions $U$ and $J$ 
for five-fold degenerate systems.

Let us first sum up all contributions where $U$ alone is relevant, and
connect these to LSDA deficiencies. $U$ dominates the strength of 
correlations. Due to $U$, electrons loose considerably more kinetic or
band energy than would be expected for a homogeneous electron gas
approximation. 
This can be seen in the Compton scattering. 
From the latter, one gets the impression that a treatment based on homogeneous
electron gas ideas must miss most of $U$.

The next topic are binding energies, equilibrum distances and 
magneto-volume effect.
All transition metals have LSDA binding energies that are too large - actually
almost exactly by the amount which is removed by the residual $U$.
Also the equilibrium distances are too short.
The LSDA magneto-volume effect is always too large - in part, it can be
corrected by effects of the residual $U$ \cite{soh, kai}.

Finally, the interaction $U$ causes sensitivities to charge  anisotropies. 
This 
is relevant for the anisotropic exchange splitting
of $Ni$. Due to its effect on the Fermi surface, it is 
basically a ground state property. LSDA lacks this splitting.

Consequently one may safely conclude that LSDA misses all $U$-contributions,
or more cautiously expressed, it reduces $U$ to $J$ (this 
would avoid attractive interactions).
This finding, however raises a problem. $U$ is very important for the
Stoner parameter and for the magnetic moment. Roughly half of the weight
in the latter comes from $U$. Consequently,  
a second error in connection with $J$ must occur.

$J$ leaves a direct imprint only on a single feature, namely
the $m$ dependence of $I(m)$. Due to the inclusion of spin correlations,
$I(0)$ is typically reduced by 10-20 percent in comparison to $I(m)$.
Sadly, other contributions (in particular correlation effects due to $U$)
cause a magnetic moment dependence, too. There is a single exception:
$Co$. LSDA shows no feature like this, but on the
other hand, there is not yet experimental evidence for this effect.

It is worth to investigate somewhat more how $I_{LSDA}$ 
is obtained from LSDA calculations.
As mentioned before, whenever an LSDA calculation is made, even for atoms, 
$I_{LSDA}$ has the same value. In the atomic case,
LSDA is assumed to describe the Hund's rule ground state, and
LDA is usually assumed to describe an average over all possible atomic
states. Consequently, $I_{LSDA}$ must describe just the atomic exchange $J$,
and it must do so in a mean-field approximation - this implies
that no correction is possible without
broken symmetry. In this respect it must behave exactly like a 
HF-theory, or like the incorrect SCF-approximation of our model Hamiltonian.

Only if LSDA behaves this way, can one understand why it obtains
the correct moment for $Fe$ \cite{soh,soh1}. 
$I$ depends roughly to equal parts on $U$ and $J$. 
Correlations reduce the effects
of these terms 
by 40 percent from the SCF limit. 
As a consequence, the error involved in skipping $U$ 
(or in reducing it to $J$),
is almost exactly compensated by the error in not correlating $J$. For
magnetism in general, this compensation works only for 5-fold
degeneracy. The error on the $U$-side is considerably larger than
the error on the $J$-side for a single-band system. Consequently
the LSDA must and usually does underestimate magnetism in general.

This latter deficiency is actually known, as the popularity 
of more recent '$LDA$ plus $U$'-approximations demonstrates.
Here, a local interaction is added to boost magnetism.
However, all attempts to generate a kind of
compensation within the DF-framework to upkeep the correct magnetic moment
of $Fe$ have failed. They had to fail because the second error, the one
for $J$
is of a different origin and thus independent.
Even worse, these two are not the only LSDA-errors in the transition
metal context as will be demonstrated
in the next section.

These findings also call a particular field of LSDA-applications into 
question,  namely all so called ab-initio
disordered local moment calculations with which one has
attempted to correct the false LSDA Stoner-theory results.
As just derived, these
LSDA disordered local moments 
are nothing but the false and inadequate model disordered local moment
approximations for the Hubbard model. The local degrees of
freedom generated by either method
are artefacts of the approximation and have no connection
to reality.
 
The only meaningful extensions of LDA-schemes are like the ones
we had made for the first time a quarter century ago \cite{st,os} 
and like those that are now made in 
connection with DMFT-applications \cite{li2}: to condense the LDA
results into a tight-binding Hamiltonian, to connect it to a
local model interaction and to perform a careful correlation treatment 
of the latter. 

A problem arises with the charge distributions
in the ground state of these models. It is often, but not always, a good choice
to freeze the charge distribution of this state to the
one of the LDA-input. Counter-examples are the changes
in the Fermi surface of $Ni$ that would not show up this way, or the 
inverse magneto-volume
effect of $Ni$ that arises from anisotropic exchange contributions
that cause a charge transfer from the $4s,4p$ to
the $3d$-orbitals. There are also systems like the 
high $T_c$-superconductors where
the LDA-charge distribution is wrong \cite{s98,st02}. As will be shown next, 
the case of $Fe$ is another example where it does not pay to
stay close to LSDA-results even for the charge distribution.

\section{Ab-initio Correlation Calculations for $Fe$}

As mentioned before, the model calculations using the LA are
only a special application of the original ab-initio scheme.
Here, first results of an
ab-initio calculation for non-magnetic $Fe$ will be
presented in order to contribute to the resolution of the open problem
of $T_c$. We adress non-magnetic Fe, because
we assume that the ferromagnetic state is rather well reproduced in LSDA. The
moment is correct, and also the Fermi surface seems to be in agreement with
experiment. The deficiences in the description of the
magnetic phase transition might instead be connected to the
non-magnetic ground state. 

Like the non-magnetic
HF-ground state and the LDA-ground state, the correlated
non-magnetic ground state is theoretically well-defined. The only possible
problem in the latter case might be that the added correlations allow
long-range ferromagnetic patterns, and that in the approximation used the
calculations turn instable.
We proceeded only up to second nearest neighbor corrections for the case of 
$Fe$, and found no instability up to this range. 

The details of the calculation will be given elsewhere \cite{sun}.
It should just be mentioned that for the HF- calculation
and the parallel LDA-calculation the program Crystal was 
used \cite{crys}.
The basis set quality was of double-zeta quality, and better for the
$3d$-electrons, and the computation was performed at the experimental
lattice constant.
A first correlation calculation starting from the LDA-ground state
single determinant worked fine. Although the calculation was performed with the
full and unscreened interaction, the electrons in the 9 valence orbitals
screened each other perfectly. Correlations were as weak as for the
model calculation with screened interaction. Note that this time
for the 9 fluctuating channels, 45 correlation channels were available.

From the $3d$-correlation patterns we could also obtain an estimate on
the effective local interactions, as mentioned before. 
This turned out to be $3 \pm 1$eV,
and was in reasonable agreement with the interaction needed for the
Hubbard model treatment.

The LDA-charge distribution was analysed using the LA. Within this
scheme, precise atomic orbitals are required for correlation purposes.
A method had been developed to unambiguously obtain these from the
single-particle density matrix $P_{ij}(l,l')$ \cite{gps}. For a similar
application, see refs. \cite{s98,st02}.
\begin{table}[hbtp]
\begin{center}
\begin{tabular}{|l|cc|cc|}
 \hline                                                                        
Orbital& HF\ \ \  &LA &LDA \ \ &LDA(tbf)\\
\hline
$4s$ &0.272&0.273&0.288&0.29\\
$4p$ &0.203&0.173&0.163&0.06\\
$3d(t_{2g})$ &0.966&0.715&0.678&0.73\\
$3d(e_g)$ &0.100&0.542&0.611&0.66\\
\hline
$\Delta$&0.866&0.173&0.067&0.07\\
\hline
\end{tabular}                                                                
\end{center}
\caption{\protect Charge distributions for bcc non magnetic $Fe$.
Given are values for the HF, LA, and LDA calculations. Also
values of a LDA tight binding fit \cite{papa} are included.}
\label{table3}
\end{table}                   
The two right columns of table \ref{table3} contain our charge analysis
in comparison to a standard tight-binding fit \cite{papa}. 
There is good agreement,
except a small charge transfer from the $4p$-orbitals into the $3d$-orbitals
in the case of the fit. We assume that this occurs because for the fit, 
the completely empty $4p$-bands needed to be included which
probably hybridize with the $4d$-bands. This apparently has an effect on 
the resulting charge distribution.
In our numerical determination, only the occupied part of the bands was 
of relevance. The occupation anisotropy $\Delta=n_{t_{2g}}-n_{e_g}$
which is most relevant comes out the same. 

When performing the HF-calculation, a very different charge distribution 
was  obtained. The $4s$- and $4p$-orbital occupations did not change but
a complete charge re-arrangement occured for the $3d$-orbitals.
The $t_{2g}$-orbitals were almost completely filled, and the $e_g$-orbitals
almost empty. This charge distribution is definitely incorect. It would never
deliver the required sizeable binding energy contributions of the $3d$-bands.
The values for the true ground state are also given. In particular $\Delta$
is considerably closer to the LDA-values. 

This charge transfer represented by $\Delta$ is originally not 
connected to on-site interactions. 
Neither did our correlation calculation based on the LDA-ground state
show instabilities toward a charge transfer, nor had the earlier
model calculations given any hint for such a behaviour.

Rather, this charge transfer is due to 
a quantity that has been almost completely disregarded
in the past: the
non-local exchange. The long-range exchange contributions per site are 
formally of the form
\begin{equation}
\Delta E_{exch}= -\sum_{ijl}V(i,0;j,l)P_{ij}(0,l)^2 .
\end{equation}
Here, V is the Coulomb interaction term between orbital $i$ on
site 0 and orbital $j$ on site $l$, and $P_{ij}(0,l)$ the corresponding
density matrix that was introduced above.
For the density matrix of a single-determinant state, the 
following sum rule applies
\begin{equation}
\sum_{jl}P_{ij}(0,l)^2=n_i.
\end{equation}
 As a consequence, delocalization pushes weight from neighbor terms
into longer-range terms and costs considerable exchange energy.
The amount is related to the size of long-range fluctuations and depends
strongly on the density of states $n(E_F)$. The latter, and even more
the peak structure around it is very large for non-magnetic $Fe$, and is
extremely
costly in exchange energy. The peak structure is formed by antibinding
$e_g$- and $t_{2g}$-orbitals. For the only non-local contribution to the
LDA-calculation, the kinetic energy, this peak is apparently irrelevant, 
but adding
only a small part of the
non-local exchange immediately starts to separate the different contributions.
The $e_g$-orbitals are pushed up, and the $t_{2g}$-orbitals are
pushed down. When computed using the LDA, then the big charge
transfer towards the LA-ground state costs 
less than 0.1eV per atom in energy. This is
negligibly small in comparison to the binding energy and still smaller than
the magnetization energy. However, 1.5eV are gained from 
the full exchange, and 
0.3eV remain when the latter is screened. The ground state charge
distribution is then further influenced by 
the strong spin correlations between the $e_g$-electrons forming 
when these reach half-filling.

\begin{figure}[h,b,t]
{\centerline{\includegraphics[width=8cm,clip,angle=0]{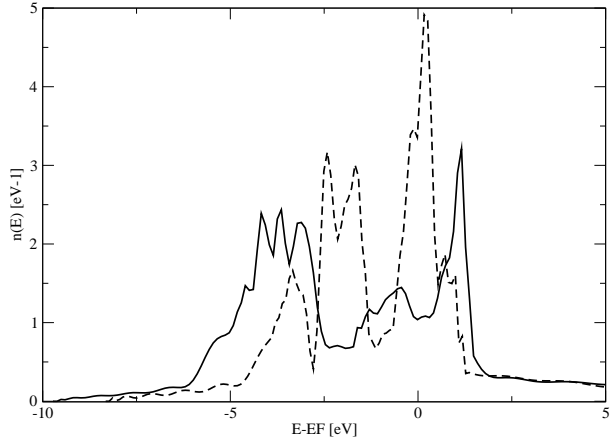}}}
\caption{Total density of states $n(E)$ for bcc nonmagnetic$Fe$,
obtained in LDA (dotted line) and by the LA (full line). $E_F$
is set to zero for both cases.}
\label{fig11}
\end{figure}

The most relevant quantity in our context is the resulting density of states.
The LDA- total density of states
(summed over spins) and the one of the LA-ground state are given in Fig. 
\ref{fig11}. As can be seen, the LDA-peak close to the Fermi energy 
splits and is shifted to both sides of it.
$n(e_F)$ is reduced from $3.2 eV^{-1}$ to $1.1 eV^{-1}$.
It becomes so small that the Stoner criterion is no longer fulfilled.
Consequently, there is a true metastable non-magnetic bcc ground state
for Fe, and the magnetic phase transition is of first order. Thus it is
no wonder that there has been no chance to obtain a reasonable transition
temperature starting from the unphysical non-magnetic ground state of the LDA.
The instability of the LDA-state is such that already an admixture of
only 5 percent of non-local exchange is sufficient to reduce the density of 
state at the Fermi energy by half. 

These non-local exchange effects create a general trend towards weak 
localization.
In $Fe$, they can act without symmetry breaking, but for single-band systems
they would enter mostly via a symmetry-lowering 
charge-density-wave instability.

The LA-density of state indicates a widening of the $3d$-bands by
1.0eV. It has been  obtained from a single-particle calculation with such a
fraction of the non-local exchange added that the correct
charge distribution was reproduced. Neither this density of state 
nor the original 
LDA-density of state contain any mass enhancements due to further
many-body contributions. The latter should amount to 15-20 percent
and shrink the band widths accordingly. 

Sadly, no experimental informations about the density of states 
above the magnetic transition temperature are available. They might
immediately verify our results since these differ strongly from LDA,
and also from the ferromagnetic result, obtained either
experimentally or in LSDA. There is
no strong peak just below the Fermi-energy either, as it originates from the
majority states in the ferromagnetic case. 

There is an experiment, though, that can be explained by these new
results: the measurements of the unoccupied Fe energy bands \cite{kir}.

\begin{figure}[h,b,t]
{\centerline{\includegraphics[width=8cm,clip,angle=0]{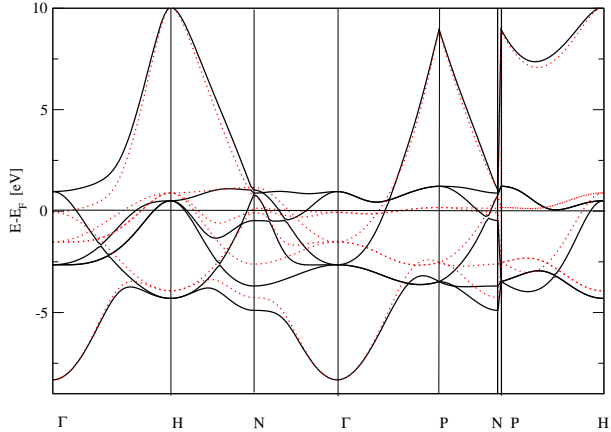}}}
\caption{Energy bands of non magnetic bcc Fe from the LDA calculation
(dots) and from the LA calculation (full lines). The occupied band width
of the latter bands is renomalized to the one of the LDA bands.}
\label{fig14}
\end{figure}

For this purpose, the energy bands of the LDA and of the LA ground state
are presented in Fig. \ref{fig14}.

The full lines represent the LA case, and the dotted bands the LDA case.
This time, the LA bands have been renormalized so that the occupied width
matches the LDA case, in order to facilitate comparison. The renormalization
factor is 0.8, and may be deduced from Fig \ref{fig11}. As can be seen, 
non-local exchange has pushed the bands with $e_g$-character around $E_F$
roughly 1 eV above $E_F$, while bands with $t_{2g}$-character are a 
little lowered (around the H and N points). This explains the changes in the
density of states. 

It is of interest to compare the energy bands of the
true non magnetic state with the ones of the ferromagnetic ground state.
The latter can be derived from the LDA bands by either an upward shift
by 1 eV for the minority bands or a downward shift by the same amount for
the majority bands. As a consequence, the unoccupied minority bands are almost
exactly where the corresponding non-magnetic bands are. An exception is
the H point where minority and majority bands enclose 
the non-magnetic band asymmetrically.

With temperature increasing toward $T_c$, the unoccupied minority bands
are therefore expected to stay in place while the bands around the H points
are shifted. This is exactly what has been measured. The unoccupied minority 
and majority bands around the H point are
shifted towards a center position, the former slightly more.
The minority
bands measured in the vicinity of the P point and in-between N and P points
did not shift at all, however \cite{kir}.
These experiments fully comfirm our results for the
non-magnetic ground state. 

So far, the experimental results have  been 
interpreted differently. It has been speculated that parts of the 
magnetic order 
do not break down when $T_c$ is reached, and that certain directions
in momentum space might keep a magnetic memory for some unknown reasons.
It would be worthwhile to re-analyse the experimental data and also to 
extract how the
majority bands become emptied and jump towards the true non-magnetic bands.

The proper computation of
the first order phase transition remains a topic
to be addressed in the future. On the model level,
this would 
imply to include
at least neighbor interactions, and is definitely
out of the range for the  present DMFT. Further speculations on
giant spin fluctuation theories should be put on hold, however,
because it is very probable that they will never be needed for $Fe$,
either.

We do not expect that non-local exchange contributions play a
role for the ferromagnetic ground state of $Fe$, except for a certain band
widening as has been discussed above (this is also why we have rescaled the
LA bands to the LDA bands in Fig. \ref{fig14}).
$n(E_F)$ is already rather low for the ferromagnetic case. The role of 
non-local exchange  contributions
shrinks further when dealing with more strongly occupied bands as
for $Co$ and $Ni$. Also, there is no peak in the  density of states
at the Fermi energy in these systems that might be removed.

On the other hand, we expect that these non-local exchange contributions
play a significant role for the $Fe$-compounds. It is known that the LSDA
usually does not treat these correctly, 
but it does so without an apparent systematic trend.
The $Fe-O$-compounds come out more magnetic in reality than using the 
LSDA, while
for the $Fe-Al$-compounds the inverse holds true. In all these cases, there
is a sizable crystal field that $Fe$ itself lacks. This would make
these compounds even more susceptible to a charge redistribution
and accompanying weak localization than $Fe$ itself.

There is not only a competition between non-magnetic
and magnetic states for the transition metals, but also one between
more or less weakly localized states. While the interactions behind
the former competition are $U$ and $J$, the ones behind the latter
are $V$.

\section{Conclusion}
                              \label{sec:conc}

The aim of this work has been to present theoretical
achievements in the computation of multi-band Hubbard models
for transition metals. Luckily, this application turned out 
to be sufficiently remote from any kind of Mott-Hubbard
transition so that currently well-controlled
weak-correlation expansions could be used. As a consequence
we can be certain that a model
and not a deficiency in the treatment is responsible if we 
do not match experiment. 
We could exclude a strong-correlation scenario for a set of
experimental findings, but the strongest argument against this scenario
is the not fully magnetized ground state of $Fe$.

This weak-correlation scenario also made it possible to
connect with LSDA-calculations, and
to analyse and explain a set of errors, even failures, 
of the latter method.

Our findings indicate the seeming success of LSDA for
the magnetism of the transition metals
is basically due to a chance compensation of two big errors: an
overscreening of $U$ on the one hand, and a mishandling of the
atomic exchange interaction $J$ in a mean-field approximation on the other.

We have shown that these model interactions are of very similar size when
obtained from fits to different experiments,
and that it is even possible to understand the trend among these
values for different transition metals. This indicates that 
these model interactions are more meaningful quantities
 than just fitting parameters.
We have also given an
indication of how these parameters will, in the future, be directly computed
from ab-initio correlation calculations using the LA.

On the model level, a set of different properties could be understood.
Also with the help of more extensive CI-calculations \cite{web1},
and in particular thanks to quasiparticle calculations using the 
DMFT  \cite{li2},
a basic understanding has finally been reached of the problems that had been
raised half a century ago \cite{sto,vv,wo,33,he}, and that 
had been mostly put aside the last forty years as far as 
the DF-approach has been concerned.

We now know that the magnetic phase transition of the transition metals
can basically be described
following the ideas of Stoner and Wohlfarth, and that the electrons
stay almost completely delocalized. We also know that
orbital degeneracy plays the most important role in this context,
and that the atomic exchange interaction is of great importance.

A very surprising twist is that for the magnetism
of $Fe$, also residual non-local
interactions seem to be relevant. There has been earlier evidence 
that they are 
present in $Fe$, originating from trends in the
dependence of $U$ on the individual atoms, and from matches and
mismatches of theoretical and experimental photoemission results.
These non-local interaction contributions lead to big changes
of the LDA-results for the non-magnetic ground state, although
those latter results were and still are the basis of a satisfying
description of the ferromagnetic ground state.

All these findings demonstrate how
important it is to understand and to treat the interactions
of the electrons directly. It is an advantage of the LA that the
correlation treatment does not disappear within the black box of a
numerical program but that all different correlation details can
be directly obtained and understood. One aim 
of this contribution was to present, with the help of the LA, 
a lucid and understandable
decomposition of  the complex many-body world.


\begin{thebibliography}{99}

\bibitem{LDA1}                                  
                                
 P. Hohenberg and W. Kohn, 
Inhomogeneous electron gas,
Phys. Rev. {\bf 136}, 864 (1964). 


\bibitem{LDA2}                                                                   
 W. Kohn and L.J. Sham, 
Self consistent equations including exchange and correlation effects,
Phys. Rev. {\bf 140}, 1133 (1965).                         

\bibitem{6} O. Gunnarsson and B.I. Lundqvist, 
Exchange and correlation in atoms, molecules, and solids by the 
spin-density-functional formalism,
Phys.Rev.{\bf B13}, 4274 (1976).


\bibitem{15} U.K. Poulson, J. Kollar, and O.K. Anderson, 
Magnetic and cohesive properties from canonical bands,
J. Phys. {\bf F6}, L 241 (1976). 

\bibitem{16} J. F. Janak and A. R. Williams, 
Giant internal magnetic pressure and compressibility anomalies,
Phys. Rev. {\bf B14}, 4199 (1976).

\bibitem{17}J. Callaway in Physics of Transition Metals (Inst. of Physics,
Bristol, 1981).

\bibitem{38} J. Friedel in The Physics of Metals (Cambridge U. Press, 
Cambridge, 1969).

\bibitem{18} V. L. Moruzzi, J. F. Janak, and A. R. Williams, Calculated
Electronic Properties of Metals (Pergamon, New York, 1978).


\bibitem{stol3}   G. Stollhoff and P. Fulde,
On the computation of electronic correlation energies within the 
local approach,
J. Chem. Phys. {\bf 73}(9), 4548 (1980).

\bibitem{stoln}   G. Stollhoff,
The local ansatz extended,
                  J. Chem. Phys. {\bf 105}, 227 (1996).

\bibitem{li} A. Heilingbrunner and G. Stollhoff, 
Ab-initio correlation calculation for metallic lithium,
J. Chem. Phys. {\bf 99}, 6799 (1993).

\bibitem{s98}
G. Stollhoff, 
Electrons in high-$T_c$ compounds: ab-initio correlation results,
Phys. Rev. B {\bf 58}, 9826 (1998).

\bibitem{sun}
G. Stollhoff, unpublished.

\bibitem{11}L. Kleinmann and K. Mednick, 
Self-energy contributions to the electronic structure of Ni,
Phys. Rev. {\bf B24}, 6880 (1981).

\bibitem{mor1} T. Moriya (ed.), Electron Correlation and Magnetism in
Narrow Band systems, Springer Ser. Solid-State Sci., Vol 29
(Springer, Berlin 1981).

\bibitem{st}
G. Stollhoff and P. Thalmeier, Variational Treatment of Electronic
Correlations in $d$-Band Metals,
Z.Physik B {\bf 43}, 13 (1981).

\bibitem{os}
A. M. Ole\'s and G. Stollhoff, Correlation effects in ferromagnetism
transition metals,
Phys. Rev. B {\bf 29}, 314 (1984).

\bibitem{sto} E. C. Stoner, Proc. R. Soc. London, 
Collective electron Ferromagnetism
Ser. A {\bf 165}, 372 (1938).
\bibitem{wo} E. P. Wohlfarth, Proc. R. Soc. London,
Collective electron Ferromagnetism III Nickel and nickel-copper alloys,
Ser. A {\bf 195}, 434 (1949).

\bibitem{gunn1} O. Gunnarsson, 
Band model for magnetism of transition metals in the spin-density-functional
formalism,
J. Phys. F {\bf 6}, 587 (1976).



\bibitem{os85} A. M. Oles, G. Stollhoff, 
Correlation Corrections for the Stoner Parameter of Ferromagnetic 
Transition Metals,
Europhys. Lett. {\bf 5}, 175 (1988). 


\bibitem{soh}
G. Stollhoff, A. M. Ole\'s, and V. Heine,
Stoner exchange interaction in transition metals,
Phys. Rev. B {\bf 41}, 7028 (1990).


\bibitem{soh1}
G. Stollhoff, A. M. Ole\'s, and V. Heine,
Comment on  Relationship between the Coulomb Integral $U$ and the
Stoner Parameter $I$,
Phys. Rev. Lett. {\bf 76}, 855 (1996).

\bibitem{mor2} T. Moriya, Spin fluctuations in Itinerant Electron Magnetism,
Springer Ser. Solid-State Sci., Vol 56
(Springer, Berlin 1985).



\bibitem{35} F. Kajzar and J. Friedel, 
Role of the intraatomic Coulomb correlation on the energy of cohesion in narrow band metals,
J. Physique {\bf 39}, 397 (1978).

\bibitem{36} G. Treglia, F. Ducastelle, and D. Spanjaard,  
Perturbation treatment of correlations in transition metals,
J. Physique {\bf 41}, 281 (1980).

\bibitem{62} J. Kanamori, 
Electron correlation and ferromagnetism in transition metals,
Progr. Theor. Phys. {\bf 30}, 275 (1963).

\bibitem{58} A. Liebsch, 
Effect of self-energy on the valence-band photoemission spectra of Ni,
Phys. Rev. Lett. {\bf 43}, 1431 (1979).

\bibitem{59} A. Liebsch, 
Ni $d$-band self-energy beyound the low-density limit,
Phys. Rev. {\bf B23}, 5203 (1981).


\bibitem{60}D. Penn, 
Effect of bound hole pairs on the $d$-band photoemission spectrum of Ni,
Phys. Rev. Lett.{\bf 42}, 921 (1979).

\bibitem{61} L. C. Davis and  L. A. Feldkamp, 
Quasi-particle energies and two-hole xps satellite in Ni,
Solid State Commun. {\bf 34}, 141 (1980).

\bibitem{web} J. B\"unemann, W. Weber, and F. Gebhard, 
Multiband Gutzwiller wave functions for general on-site interactions
Phys. Rev. B {\bf57}, 6896 (1998).

\bibitem{web1} J. B\"unemann et al.,
Atomic correlations initinerant ferromagnets: Quasi-particle bands of Ni,
Europhys. Lett. {\bf 61}, 667 (2003).

\bibitem{li2}A. Lichtenstein, M. I. Katsnelson, and G. Kotliar,
Finite-temperature magnetism of transition metals:
an ab initio dynamical mean-field theory,
Phys. Rev. Lett. {\bf 87}, 067205 (2001).

\bibitem{jep}O. K. Andersen and O, Jepsen, 
Advances in the theory of one-electron energy states,
Physica   {\bf B 91}, 317 (1977).

\bibitem{33} M. C. Gutzwiller, 
Corredlation of electrons in a narrow $s$ band
Phys. Rev. {\bf 137}, 1726 (1965).

\bibitem{jastr} R. Jastrow,
Many-Body problem with strong forces,
Phys. Rev {\bf 98}, 1479 (1955).

\bibitem{cc1}F. Coester, Nucl. Phys. {\bf 7},421 (1958).

\bibitem{cc2}F. Coester and H. K\"ummel, Nucl. Phys. {\bf 17}, 477 (1960).

\bibitem{cc3}J. Cizek, 
On the correlation problem in Atomic and Molecular Physics
J. Chem. Phys.{\bf 45}, 4256 (1966).

\bibitem{cc4}J. Cizek, Adv. Chem. Phys. {\bf 14}, 35 (1969).



\bibitem{vo} D. Vollhardt et al., 
Metallic ferromagnetism: progress in our understanding of an old 
strong-coupling problem,
Advances in Solid State Physics, 
Vol.38, 383 (1999).



\bibitem{stohei} G. Stollhoff, and A. Heilingbrunner, 
Coupled-cluster equations for the local ansatz,
Z. Physik {\bf B83},85 (1991).


\bibitem{un} P. Unger, J. Igarashi, and P. Fulde, 
Electronic excitations in $3d$ transition metals,
Phys. Rev. 
B{\bf 50}, 10485 (1994).

\bibitem{lic} M. I. Katsnelson and A. I. Lichtenstein,
LDA++approach to the electronic structure of magnets: correlation
effects in Fe,
J. Phys. Cond. Matter {\bf 11}, 1037 (1999).

\bibitem{s85} G. Stollhoff, 
Magnetic correlations in the paramagnetic state of $3d$-transition metals,
J. Magn. Magn. Mat. {\bf 54-57}, 1043 (1986).

\bibitem{52}P. J. Brown, D. Deportes,D. Givord, and K. R. A. Ziebeck,
Paramagnetic scattering studies of the short-range order above $T_c$
in $3d$ transition metals compounds and pure Fe,
J. Appl. Phys. {\bf 53}, 1973 (1982).



\bibitem{cstei} M. M. Steiner, R. C. Albers, and L. J. Sham,
Quasiparticle properties of Fe, Co, and Ni, 
Phys. Rev. {\bf B 45}, 13272 (1992).                                                                                


\bibitem{54}V. Korenman, J. L. Murray, and R. E. Prange, 
Local band theory of itinerant ferromagnetism. I Fermi-liquid theory,
Phys. Rev. {\bf B16},
4032 (1977)

\bibitem{55} H. Capellmann, 
Ferromagnetism and strong correlations in metals,
J. Phys. {\bf F 4}, 1466 (1974).

\bibitem{53} H. Capellmann, 
Theory of itinerant ferromagnetism in the $3d$-transition metals,
Z. Physik {\bf B 34}, 29 (1979).


\bibitem{c1}                                                                   
G. E. Bauer and J. R. Schneider, 
Electron correlation effect in the momentum density of copper metal, 
Phys. Rev.{\bf B 31}, 681 (1985).                                                                        

\bibitem{c2}
A. J. Rollason, M. Cooper, and R. S. Holt, 
Directional Compton profiles and the electron density distribution in V,
Philos. Mag.{\bf  B 47}, 51 (1983).

\bibitem{c3}
S. Wakoh and M. Matsumoto, 
Correlation correction on Compton profiles of V and Cr,
J. Phys. Cond. Matter {\bf 2}, 797 (1990).

\bibitem{c4}
D. L. Anastassopoulos, G. D. Priftis, N. I. Papanicolaou,
N. C. Bacalis, and D. A. Papaconstantopoulos,
Calculation of the electron moment density and Compton scattering measurements
for Ni, J. Phys. Cond. Matter {\bf 3}, 1099 (1991).

\bibitem{c5}
A. J. Rollason, R. S. Holt, and M. J. Cooper, 
Directional Compton profiles and the electron density distribution
in Fe,
J. Phys. {\bf F 13}, 1807 (1983).

\bibitem{c6}
V. Sundararajan, D. G. Kanhere, and R. M. Singru, 
Calculation of spin-dependent momentum distributions in Fe:
Ccompton scattering and positron annihilation,
J. Phys. Cond.Matter {\bf 3}, 1113 (1991).

\bibitem{s952} G. Stollhoff, 
Compton scattering and Electron Correlations in the Transition Metals,
Europhys. Lett. {\bf 29}, 463 (1995).

\bibitem{him1}D. E. Eastman, F. J. Himpsel, and J. A. Knapp,
Experimental band structure and temperature-dependent magnetic
exchange splitting of Ni using angle-resolved photoemission
Phys. Rev. Lett. {\bf 40}, 1514 (1978).

\bibitem{plu}W. Eberhardt and E. W. Plummer,
Angle-resolved photoemiision determination of the band structure
and multielectron excitations in Ni,
Phys. Rev. {\bf B 21}, 3245 (1980).

\bibitem{him}D. E. Eastman, F. J. Himpsel, and J. A. Knapp,
Experimental Exchange-split energy-band dispersions for Fe, Co, Ni,
Phys. Rev. Lett. {\bf 44}, 95 (1980).



\bibitem{cooke} J. F. Cooke, J. W. Lynn, and H. L. Davis,
Calculations of the dynamic susceptibility of Ni and Fe,
Phys. Rev. {\bf B 21}, 4118 (1980).

\bibitem{saw}
D. van der Marel and G. A. Sawatzky,
Electron-elecron interaction and localization in $d$ and $f$
transition metals, 
Phys. Rev. B {\bf 37}, 10674 (1988).

\bibitem{s951}
G. Stollhoff, 
From ab-initio Calculations to the Hubbard Model and beyound,
Europhys. Lett. {\bf 30}, 99 (1995).

\bibitem{cgunn} V. I. Anisimov and O. Gunnarsson, 
Density -functional calculation of effectrive Coulomb interactions
in metals,
Phys. Rev. B {\bf 43}, 7570 (1991).


\bibitem{gudra}V. Drchal, O. Gunnarsson and O. Jepsen,
Effrective Coulomb interaction in metallic  $3d$-systems:
comparison of theory and experiment for Mn in Ag
Phys. Rev. B {\bf 44}, 3518 (1991).


\bibitem{ks}G. K\"onig and G. Stollhoff, 
Why polyacetylene dimerizes: results of ab initio computations,
Phys. Rev. Lett. {\bf 65}, 1239 (1990).

\bibitem{st02}G. Stollhoff, 
Electron correlations in the high-$T_c$ compounds,
Cimtec 2002 Proc. Forum on New Materials
Vol.3, pp229 (2002).

\bibitem{kai} A. B. Kaiser, A. M. Oles, and G. Stollhoff,
Volume dependence of the Stoner Parameter in Transition Metals,
Phys. Scr. {\bf 37}, 935 (1988).


\bibitem{crys}V.R. Saunders et al., Crystal98 User's Manual, University of 
Torino (1998).

\bibitem{gps} R. Pardon, J. Gr\"afenstein, and G. Stollhoff,
Ab initio ground-state correlation calculations for semiconductors
with the local ansatz, 
Phys. Rev. B{\bf 51}, 10556 (1995).

\bibitem{papa} D. A. Papaconstantopoulos, The band structure of
Elemental solids (Plenmu, New York,1968).                                                                               
\bibitem{kir}J. Kirschner, M. Gloebl, V. Dose, and H. Scheidt,
Wave-vector-dependent temperature behaviour of empty bands in 
ferromagnetic Iron,
Phys. Rev. Lett {\bf 53}, 612 (1984). 


\bibitem{vv} J. H. van Vleck, Rev. Mod. Phys. {\bf 17}, 27 (1945).


\bibitem{he} C. Herring, in Magnetism (Eds. G. Rado, H. Suhl) Vol. 4
(Academic Press,1966).


\end{thebibliography}
\end{document}